\documentclass[sn-vancouver,Numbered]{sn-jnl}

\usepackage{subcaption}
\usepackage{graphicx}%
\usepackage{multirow}%
\usepackage[compatibility=false]{caption}
\usepackage{placeins}
\usepackage{amsmath,amssymb,amsfonts}%
\usepackage{amsthm}%
\usepackage{mathrsfs}%
\usepackage[title]{appendix}%
\usepackage{xcolor}%
\usepackage{textcomp}%
\usepackage{soul}
\usepackage{xcolor}
\sethlcolor{yellow} 
\usepackage{float}
\usepackage{manyfoot}%
\usepackage{booktabs}%
\usepackage{algorithm}%

\usepackage{algorithmicx}%
\usepackage{algpseudocode}%
\usepackage{listings}%
\usepackage{xurl} 
\usepackage{hyperref}
\usepackage[hyphenbreaks]{breakurl}

\renewcommand{\appendixname}{Appendix\\}
\begin{document}
	
\title[Article Title]{Block-Fused Attention-Driven Adaptively-Pooled ResNet Model for Improved Cervical Cancer Classification}
	
	\author[1]{\fnm{Saurabh} \sur{Saini}}\email{phd2101101005@iiti.ac.in}
	\author*[1]{\fnm{Kapil} \sur{Ahuja}}\email{kahuja@iiti.ac.in}

	\author[2]{\fnm{Akshat} \sur{S. Chauhan}}\email{akshatschauhan1@gmail.com}
	
	\affil[1]{\orgdiv{Math of Data Science \& Simulation (MODSS) Lab, Computer Science \& Engineering}, \orgname{Indian Institute of Technology Indore},  \country{India}}
	
	\affil[2]{\orgdiv{Computer Science \& Engineering}, \orgname{Indian Institute of Information Technology Nagpur}, \country{India}}

\abstract{
		
		 	Cervical cancer is the second most common cancer among women and a leading cause of mortality. Many attempts have been made to develop an effective Computer Aided Diagnosis (CAD) system, however, their performance remains limited. Using pretrained ResNet-50/101/152, we propose a novel CAD system that significantly outperforms prior approaches.
		 	 
		 Our novel model has three key components. \textit{First}, we extract detailed features (color, edges, and texture) from early convolution blocks and the abstract features (shapes and objects) from later blocks, as both are equally important. This dual-level feature extraction is a new paradigm in cancer classification. \textit{Second}, a non-parametric 3D attention module is uniquely embedded within each block for feature enhancement. \textit{Third}, we design a theoretically motivated innovative adaptive pooling strategy for feature selection that applies Global Max Pooling to detailed features and Global Average Pooling to abstract features. These components form our Proposed Block-Fused Attention-Driven Adaptively-Pooled ResNet (BF-AD-AP-ResNet) model. To further strengthen learning, we introduce a Tri-Stream model, which unifies the enhanced features from three BF-AD-AP-ResNets. An SVM classifier is employed for final classification.
		 
		 We evaluate our models on two public datasets, IARC and AnnoCerv. On IARC, the base ResNets achieve an average performance of \(90.91\%\), while our model achieves an excellent performance of \( \textbf{98.63\%}\). On AnnoCerv, the base ResNets reach to \(87.68\%\), and our model improves this significantly, reaching \( \textbf{93.39\%}\). Our approach outperforms the best existing method on IARC by an average of \( \textbf{14.55\%}\). For AnnoCerv, no prior competitive works are available.  We conduct ablation studies to justify the inclusion of each component. Additionally, we introduce a novel SHAP+LIME explainability method, accurately identifying the cancerous region in $\textbf{97\%}$ of cases, ensuring model reliability for real-world use.

		}


	\maketitle

\section{Introduction} \label{sec: Introduction}
Cervical cancer is a serious health concern for women aged 15–44, ranking second in fatality after breast cancer with 340,000 deaths annually \cite{xu2017multi}. It is also the second most diagnosed cancer in women, with more than 600,000 annual cases \cite{sung2021global}. The main cause of this is the Human Papilloma Virus (HPV) transmitted through sexual contact \cite{siegel2021cancer}. 

Early screening programs can save lives since this cancer is considered incurable in the advanced stages. Screening methods include visual inspection with acetic acid (VIA), HPV test, pap smear, and colposcopy \cite{xu2017multi}. Colposcopy is a widely used method that involves taking pictures of the cervix after applying acetic acid \cite{hua2020lymph}. This helps identify the various degrees of abnormalities visible in shades of white \cite{mayrand2007human}. There is a shortage of expert healthcare professionals who can accurately classify colposcopy cancer images (into different types and Cervical Intraepithelial Neoplasia (CIN) scores). A typical CAD system consists of three primary components, namely, feature extraction, classification, and explainability.

Feature extraction can be performed using either handcrafted descriptors or deep learning descriptors \cite{masmoudi2022optimal}. Handcrafted descriptors capture selected set of features, while the deep learning descriptors capture all types of features  \cite{masmoudi2022optimal, shehab2024deep}. For colposcopy cancer image classification, all kinds of features are important \cite{asiedu2018development}, and hence, this makes deep learning a more suitable approach here.

In deep learning, Convolutional Neural Networks (CNNs) are a popular choice. They typically come with a variety of pretrained models, such as Inception, VGG16, ResNet, AlexNet, etc. \cite{kalbhor2023cervical}, which are highly reliable because they are trained on large datasets like ImageNet. Among them, ResNet variants have recently demonstrated superior accuracy in medical imaging tasks compared to other pretrained networks \cite{kanna2024deep, hasanah2023deep, yan2021multi, yuan2020application, ma2020combining}. A key advantage of ResNets is their ability to address the vanishing gradient problem. However, despite their success, ResNets still exhibit relatively low performance \cite{dash2023cervical, masmoudi2022optimal, saini2020colponet, xu2017multi}, which we aim to improve here. 

Given their proven effectiveness, ResNets form our base models. Specifically, we choose ResNet50, ResNet101, and ResNet152 because the lower numbered ResNets capture simpler patterns while the higher numbered capture more complex patterns. To improve these ResNets we add three novel components, which we discuss now. 

The \textit{first} component focuses on feature extraction. It has been shown in pretrained neural networks that earlier convolution blocks capture detailed features such as color, edges, and texture, while the later convolution blocks capture abstract features like shapes and objects. These abstract features dominate the output of the network. For colposcopy cancer image classification, detailed features are equally important. Hence, we extract features at the end of each convolution block of the ResNets. This the first attempt of its kind in any type of cancer classification.

The \textit{second} component focuses on feature enhancement. In colposcopy images, lesions are hard to see because of variation in shape/size/position of the lesions, image reflections, overlapping tissues, poor lighting, etc. We uniquely embed a non-parametric 3D attention module within each convolution block throughout the network to prioritize the most relevant features and suppress the noisy ones.

The \textit{third} component focuses on feature selection to reduce the large feature space. We introduce innovative adaptive pooling mechanism that applies max pooling to the detailed features and average pooling to the abstract features. This is because detailed features are noise-prone, hence, we aim to select the most relevant ones, while abstract features are not affected by noise, so we prefer to retain all of them. 

These three components collectively define our powerful Block-Fused Attention-Driven Adaptively-Pooled ResNet (BF-AD-AP-ResNet) model. We generalize this research idea for all three ResNets (50/101/152), resulting in BF-AD-AP-ResNet50, BF-AD-AP-ResNet101, and BF-AD-AP-ResNet152, respectively. Furthermore, we develop a Tri-Stream model to expand the richness of features. Our Proposed Tri-Stream model merges features from three Proposed BF-AD-AP-ResNets models.

After feature extraction, we need to classify the colposcopy cancer images. When the two parts are combined, the model is referred to as an end-to-end model that typically works well on medium to large-sized datasets. When feature extraction is done separately from classification, then the model is referred to as a hybrid model that works well even on small to medium-sized datasets \cite{kalbhor2023cervical, saidi2025efficient}. The publicly available datasets under consideration here (see below) are small to medium, and hence, we work with a hybrid model while using a Support Vector Machine (SVM) for classification. To evaluate our CAD system, we perform both the 5-fold and the 10-fold cross-validation. In each case, the dataset is divided into training and testing sets. The training and testing sets are normalized separately before being fed into the SVM for classification. 

There are only two publicly available high-resolution colposcopy cancer image datasets\footnote{There is also an older public dataset called the Intel \& MobileODT dataset \cite{intel-mobileodt-cervical-cancer-screening}, which is available on Kaggle. However, the image quality is low because the images were captured using mobile-level devices \cite{wu2024artificial}.}, namely, IARC \cite{IARC2024} and AnnoCerv \cite{Minciuna2023}. We perform our experiments on both the IARC and the AnnoCerv datasets. Originally, the IARC dataset contains $571$ images, and based upon a recent previous work \cite{dash2023cervical}, we apply data augmentation techniques to expand this dataset, resulting in $4725$ images. The standard ResNets achieve an average performance of $90.91\%$ on this dataset, while our best model achieves an average performance of $\textbf{98.63\%}$ on the same dataset. The best approach in literature only achieved an average performance of $84.46\%$ on this dataset. 

The AnnoCerv dataset originally contains $531$ images, which we again augment using the same approach leading to a total of $4665$ images. On this dataset,  the standard  ResNets achieve an average performance of $87.68\%$, while our best model achieves an average performance of $\textbf{93.39\%}$. On this dataset, there is no competitive approach available for comparison.

Next, We conduct an ablation study to carefully examine how each part of our model contributes to its overall performance. These parts include extracting both detailed and abstract features to capture rich information, using the attention module to enhance important features, and applying adaptive pooling to select the most relevant ones. This process helps us confirm the value of each component, understand how they work together, and improve the performance of our model.

Finally, we look at {\it explainability}. To enhance the trustworthiness and interpretability of the model outputs for healthcare experts, we apply an ensemble of explainable AI (XAI) techniques, namely, SHAP (Shapley Additive Explanations) and LIME (Local Interpretable Model-agnostic Explanations). As far as we know, this combination of XAI techniques is being done for the first time.

Cervical cancer usually starts in the cervix, hence, for a model to be trusted by healthcare experts, it should pay attention to that area. The region identified by our ensemble XAI technique on a subset of our first dataset, which contributes to the classification decision, is located around the cervix in $\textbf{97\%}$ of the cases. This shows that the decisions of our model match what healthcare experts expect. This not only proves our classification is correct but also makes the system clearer and more trusted in real medical use.

\vspace{0.5em}
\noindent To summarize, this research offers four distinct contributions:
\begin{itemize}
	
	\item We develop a novel model, BF-AD-AP-ResNet, consisting of three primary components. \textit{First} is extracting detailed and abstract features separately, which is new in cancer classification. \textit{Second} is uniquely integrating a non-parametric 3D attention module within each convolution block. \textit{Third} is innovatively applying Gobal Max Pooling (GMP) for detailed features and Global Average Pooling (GAP) for abstract features.
	
	\item We construct a Tri-Stream framework, which consolidates the features from three Proposed BF-AD-AP-ResNets.
	
	\item We exhaustively test on two publicly available datasets and achieve performance ranging from mid-nineties to the late-nineties for both datasets, which is considered to be excellent. Our results surpass all previously reported works in the literature. 
	
	\item  We also perform a ablation studies, which demonstrates the usefulness of each of the above three components.
	
	\item Among existing XAI models, it is difficult to identify the one best suited for a particular classification task. Therefore, we introduce an ensemble of SHAP and LIME to explain classification decisions, achieving above ${97\%}$ performance and making our CAD system transparent and clinically trustworthy.
	
\end{itemize}

The rest of this manuscript consists of five more sections. In Section \ref{sec: Litreture}, we review the existing literature. The designing of our model is described in Section \ref{sec: methodology}. In Section \ref{sec: Result}, we discuss the numerical results. We provide the results of ablation study of our model in Section \ref{sec: ablation}. In Section \ref{sec:xai}, we explain the decision of our CAD system using explainable AI techniques. Finally, the conclusion and the future work are given in Section \ref{sec:conclusion}.

\section{Literature review} \label{sec: Litreture} 
Table \ref{tab:survey} provides a summary of existing work on colposcopy cancer image classification. These studies performed different types of classification (2-way, 3-way, and 5-way) based on the annotations available provided in the respective datasets.

\renewcommand{\arraystretch}{1.3}
\begin{table*}[!htbp]
	
	\centering
	\caption{ Past work in classification of colposcopy images} 
	
	\vspace{5pt}
	\resizebox{0.95\textwidth}{!}{%
		
		\begin{tabular}{l|l|l|l|c|l|c|c|c}
			\hline

			\multirow{2}{*}{References}  & \multirow{1.25}{*}{Classifica- } &Type of& Extraction& Classifica-& \multirow{2}{*}{Dataset}  &\multirow{2}{*}{Sp.($\%$)}&\multirow{2}{*}{Sen.($\%$)}& \multirow{2}{*}{Acc.($\%$)} \\ 
			& tion Type  &Model&Technique&tion model&  & &  & \\  
		
			\hline

			&&&&&& &&\\
			Xu et al. \cite{xu2017multi}&  2-way &End-to-&  CaffeNet, &\textbf{--}& Private data-  & $83.40$& $ 88.30$ & $83.42$ \\
			(2017)&classifica-&End,&  & &set from NCI & && \\
			&tion&Hybrid&PHOG,&SVM,&Guanacaste&&&\\
			&&&PLAB,&AdaBoost& ($1112$ images) &&&\\ 
			&&&PLBP&&\cite{herrero1997design}&&&\\
			&&&&&&&&\\

			Saini et al. \cite{saini2020colponet}  & 2-way &End-to-&ColpoNet&\textbf{--}& Private data-  &\textbf{--}&\textbf{--}& $81.35$ \\
			(2021)&classifica-&End&&& set from  NCI    &&&\\
			
			&tion& &&& ($800$ images)&&& \\
			
			&&&&&\cite{saini2020colponet}& &&\\
			&&&&&& &&\\
			
			Yan et al. \cite{yan2021multi}  & 2-way &End-to-&ResNet18&\textbf{--}& Private data-  &$95.70$&$74.60$& $85.50$ \\
			(2021)&classifica-&End&&& set from SRRS   &&&\\
			
			&tion& &&& Hospital &&& \\
			
			&&&&&($1400$ images)&&&\\
			&&&&&\cite{yan2021multi}& &&\\
			&&&&&& &&\\

			Yuan et al. \cite{yuan2020application}  & 2-way &End-to-&ResNet50&\textbf{--}& Private data  &$82.62$&$85.38$& $84.10$ \\
			(2021)&classifica-&End&&&set ($22330$&  &&\\
			
			&tion& &&&  images) \cite{yuan2020application} &&& \\

			&&&&&& &&\\
			\hline
			
			&&&&&& &&\\
			
			Cho et al. \cite{cho2020classification}   & 5-way  & End-to- & Inception- & \textbf{--} & Private data- &\textbf{ --} & \textbf{--} & $51.70$ \\ 
			(2023) & classifica-  &End& ResNet-V2, && set ($791$)  &&&\\
			& tion &&ResNet152&&  images \cite{cho2020classification}  &&& \\

			&  & &&&   &    && \\ \hline
			
			&&&&&& &&\\
			
			Dash et al. \cite{dash2023cervical}   & 3-way  & Hybrid & Inception- & SVM & Public data- & $90.62$ & $81.24$ & $81.24$ \\ 
			(2023) & classifica-  && ResNet-V2 &&set from IARC &&&\\
			& tion &&&& $292$ images  &&& \\

			&  & &&&    \cite{dash2023cervical}   &&& \\

			\hline

	\end{tabular}}
	\label{tab:survey}
	
	\end{table*}
	
	Xu et al. \cite{xu2017multi} in $2017$ performed a 2-way (normal-abnormal) classification. They explored both the end-to-end and the hybrid approaches. For the end-to-end approach, they utilized a pretrained CaffeNet model. For hybrid, they used PLAB, PHOG, PLBP, and CaffeNet for feature extraction, and SVM and AdaBoost were used for classification. For this study, the author used a total of $1112$ images from the private dataset. The results demonstrated that the end-to-end deep learning model slightly outperformed the hybrid model, achieving an accuracy rate of $83.42\%$.
	
	Saini et al. \cite{saini2020colponet} in $2020$ also performed a 2-way classification. They used an end-to-end CNN model referred to as ColpoNet. This model was evaluated on a private dataset, which contained a total of $800$ images. They achieved an accuracy of $81.35\%$.
	
	Yan et al. \cite{yan2021multi} in $2021$ again performed a 2-way classification. They used an end-to-end pretrained CNN model, namely, ResNet18. For this study, a private dataset containing $1400$ image has been used to evaluate the performance of the model. This model also achieved an accuracy of $85.50\%$.
	
	Yuan et al. \cite{yuan2020application} in $2021$ also performed a 2-way classification. They used an end-to-end pretrained ResNet50 model. To evaluate their model, they used a private dataset of $22330$ images. The model attained an accuracy of $84.10\%$.
	
	Cho et al. \cite{cho2020classification} in $2020$ designed a model for 5-way classification. This classification was based on Cervical Intraepithelial Neoplasia (CIN) scores. Here, the images are categorized into normal, CIN1, CIN2, CIN3, and cancer. This was an end-to-end model. They used two pre-trained CNNs, namely, Inception-ResNet-V2 and ResNet152. They tested their model on a private dataset consisting of $791$ images. They showed that ResNet152 achieved higher performance than Inception-ResNet-V2, with an accuracy of $51.70\%$.
	
	Recently, Dash et al. \cite{dash2023cervical} in $2023$ conducted a 3-way classification. Here, the dataset is categorized into three types based on their Transformation Zone (TZ), namely, Type1, Type2, and Type3. This was a hybrid model, where they used a modified Inception-ResNet-V2 model to extract the features at multiple scales from the colposcopy images and merge them. They used a linear SVM for classification. They evaluated their model on $292$ images taken from the IARC dataset. Their results showed an accuracy of $81.24\%$.
	
	Most of the above works have used standard ResNets. In this work, we propose a sophisticated variant of ResNet, which has three components, separate extraction of detailed and abstract features, the use of a non-parametric 3D attention module, and an adaptive pooling technique for feature selection. This novel idea is generalized across three ResNets (50/101/152). Furthermore, by combining the features extracted from these three proposed models, we achieve a substantial performance gain over existing studies.

	\section{Model design} \label{sec: methodology}
\begin{figure*}[!h] 
\centering
\includegraphics[width=0.95\textwidth]{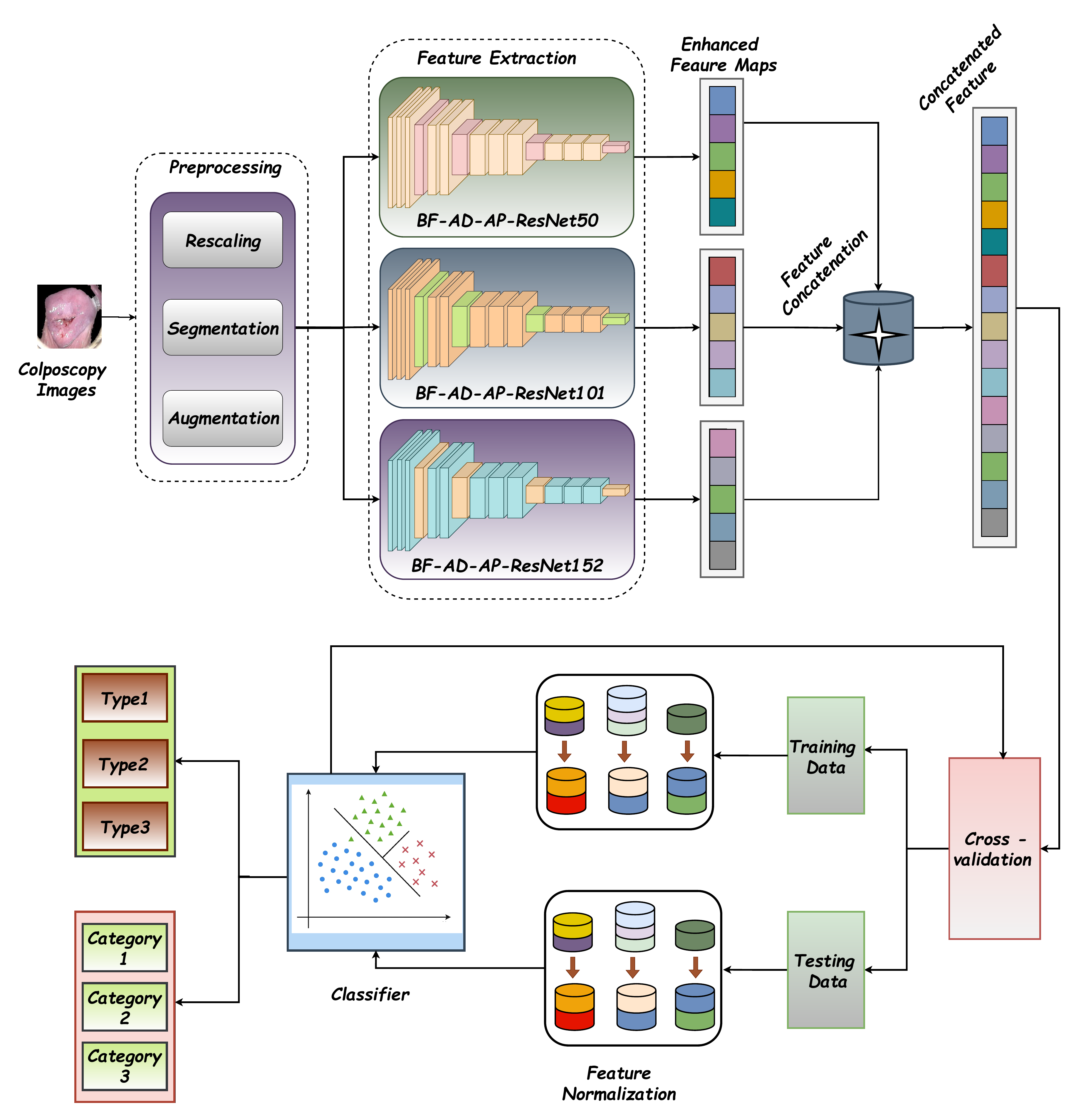}
\caption{Flow diagram of CAD system with Proposed Tri-Stream model}
\label{fig:arch_1}
\end{figure*}

	This section presents the detailed functioning of our proposed CAD system, which consists of four main components; data preprocessing, feature extraction, feature normalization, and classification. The complete setup of our proposed CAD system is shown in Fig. \ref{fig:arch_1}, and different components are discussed in the following section.
	
\subsection{Data preprocessing} 

First, we discuss the data preprocessing for the International Agency for Research on Cancer (IARC) dataset, provided by the World Health Organization (WHO) \cite{IARC2024} \footnote{This dataset is not transferrable and is meant for this research only. If anyone needs it further, then they may contact the main source.}. As discussed in literature, this dataset is categorized based on the transformation zone (TZ). The TZ refers to the region around the cervical opening where the endocervix (inner part) and ectocervix (outer part) converge (where the squamous epithelium cells replace the columnar epithelium cells). The transformation zone is typically classified into three category, namely, Type1, Type2, and Type3. For this dataset, we follow the exact same preprocessing methodology outlined in \cite{dash2023cervical}.

\begin{figure*}[!h]
\centering
\includegraphics[width=0.7\linewidth]{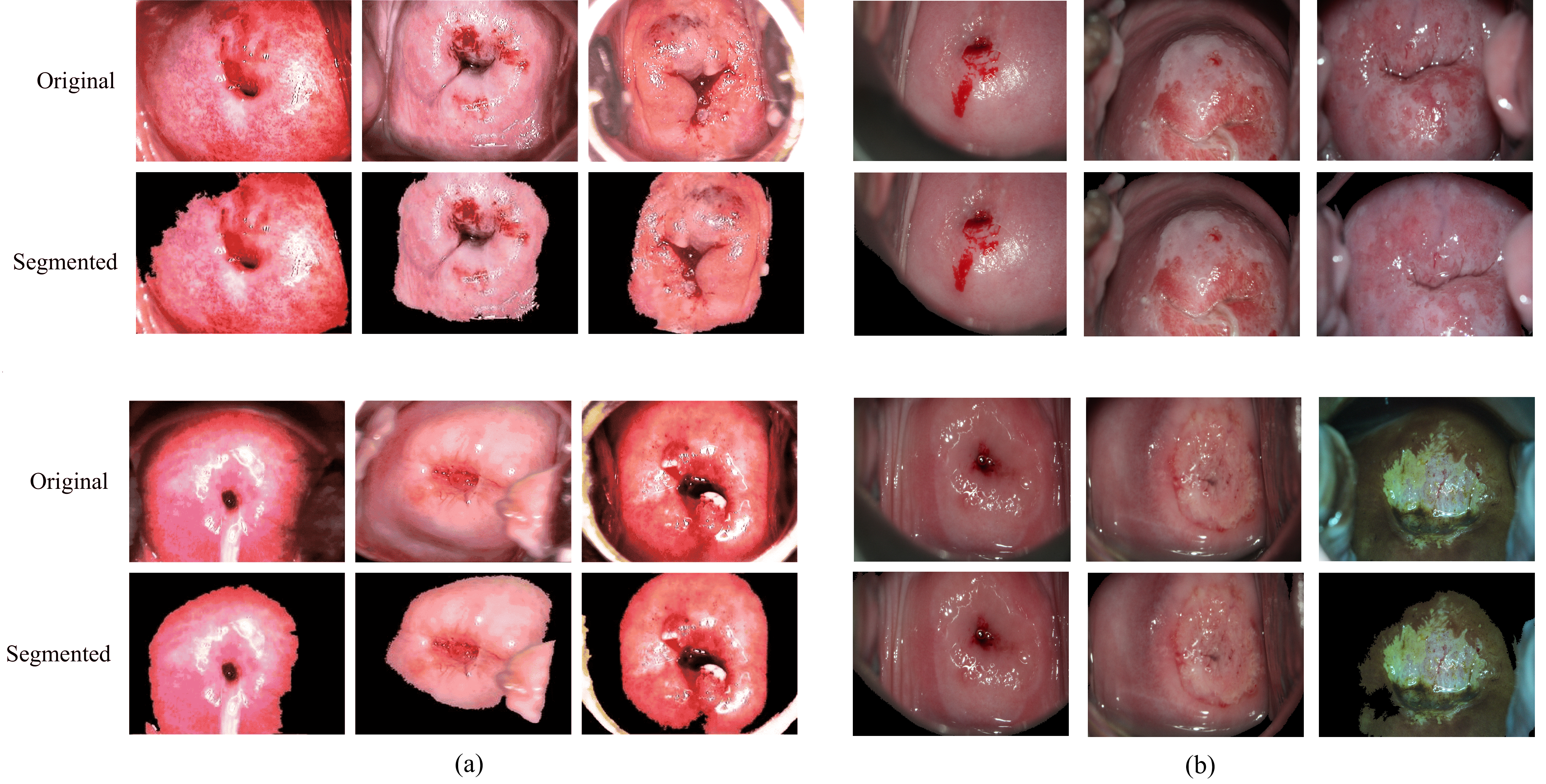}
\caption{ Rows represents the original and their segmented counterparts, column reresents the class of the image in respective manner. (a) Type1, Type2, and Types3 images of the IARC dataset \cite{IARC2024}. (b) Category1, Category2, Category3 images of the AnnoCerv dataset \cite{Minciuna2023}}
\label{fig:dataset_image}
\end{figure*}

The original colposcopy images are of $800\times600$ pixels. We rescale all images to $224\times224$ to optimize the computational resources. These images not only contain the cervical region (Region of Interest or ROI), but they also contain some artifacts that are not relevant. Therefore, we segment the images, which reduces the impact of additional artifacts beyond ROI. Some images are removed during segmentation because the cervix is not clearly visible in the original image, and the segmented result contains only irrelevant region. Fig. \ref{fig:dataset_image}(a) shows the sample of original images and their segmented counterparts for each types. 

The total number of original images and the segmented images for the types classification are given in columns 2 and 3 of Table \ref{tab:dataset_1}. As evident from this table, the data is imbalanced. To balance the types, we increase the number of images for the class with the fewest samples by a factor of 5 via rotation and flip. We match the images of the other two classes with the final number obtained above (again by rotating and flipping them). This results in 315 images for each class, as given in column 4 of Table \ref{tab:dataset_1}. To further increase the data five times, we perform random variations of contrast, brightness, rotation, and translation. This results in 1575 images for each class for a total of 4725 images, as given in columns 5 and 6 of Table \ref{tab:dataset_1}, respectively.

\renewcommand{\arraystretch}{1.1}
\begin{table*}[!h]

\centering

\caption{Dataset balancing and augmentation for the IARC dataset}
\vspace{5pt}
\resizebox{.75\textwidth}{!}{
	\begin{tabular}{c|c|c|c|c|c}
		\hline
		\multirow{3}{*}{Category} & \multirow{1.85}{*}{Original} & \multirow{1.85}{*}{Segmented}& \multirow{1.85}{*}{Rot.}  & Rand. Var.  &  \multirow{1.85}{*}{No. of}  \\ 
		& \multirow{1.75}{*}{Image}  &\multirow{1.75}{*}{Images}   & \multirow{1.85}{*}{\&  Flip} & (Cont., Bright., & \multirow{1.75}{*}{Images}  \\ 
		&  &  & &  Rot., Trans.) &   \\ \hline 
		
		Type1    & 318    & 226   & 315  & 1575  & \multirow{3}{*}{4725} \\ 
		Type2    & 106   & 63  & 315 & 1575  &     \\ 
		Type3    & 147  & 87 & 315 & 1575   &  \\ \hline
		
\end{tabular}} 
\begin{tablenotes}
	\centering
	\scriptsize
	\item Rand. Var. = Random Variation,  Bright. = Brightness, Rot. = Rotation, Cont. = Contrast, and Trans. = Translation
\end{tablenotes}

\label{tab:dataset_1}
\end{table*}

The colposcopy images given in the AnnoCerv dataset are categorized into three categories based on their scores, ranging from $0-10$. The images with scores $0-4$ fall into the first category (low-grade/CIN1), $5-6$ fall into the second category (high-grade/non-invasive cancer/CIN2+), and $7-10$ belong to the third category (high-grade/suspected invasive cancer/CIN2+) \cite{Minciuna2023}.

Next, we follow the same preprocessing steps as above for the AnnoCerv dataset. Table \ref{tab:dataset_3} lists the results of the segmentation and the augmentation and Fig. \ref{fig:dataset_image}(b) depicts the sample of original images and their segmented counterparts for each category.

\renewcommand{\arraystretch}{1.1}
\begin{table*}[!h]

\centering

\caption{Dataset balancing and augmentation for the AnnoCerv dataset}
\vspace{5pt}
\resizebox{.75\textwidth}{!}{
	\begin{tabular}{c|c|c|c|c|c}
		\hline
		\multirow{1.85}{*}{Category} & \multirow{1.85}{*}{Original} & \multirow{1.85}{*}{Segmented}& \multirow{1.85}{*}{Rot.}  & Rand. Var.  &  \multirow{1.85}{*}{No. of}  \\ 
		\multirow{1.75}{*}{Score} & \multirow{1.75}{*}{Image}  &\multirow{1.75}{*}{Images}   & \multirow{1.85}{*}{\&  Flip} & (Cont., Bright., & \multirow{1.75}{*}{Images}  \\ 
		&  &  & &  Rot., Trans.) &   \\ \hline 
		
		0-4    & 311    & 311 & 311  & 1555  & \multirow{3}{*}{4665} \\ 
		5-6    & 124   & 124  & 311 & 1555  &     \\ 
		7-10    & 96  & 96 & 311 & 1555   &  \\ \hline
		
\end{tabular}}

\label{tab:dataset_3}
\end{table*}

\subsection{Feature extraction}
This section is further divided into two subsections. The first subsection discusses the proposed Block-Fused Attention-Driven Adaptively-Pooled ResNet (BF-AD-AP-ResNet) models. The second subsection presents the Tri-Stream version of our proposed models, where features from all three models are synergistically combined to enhance the overall classification performance.

\subsubsection{Proposed Block-Fused Attention-Driven Adaptively-Pooled ResNet (BF-AD-AP-ResNet) model}

\begin{figure*}[!h] 
\centering
\includegraphics[width=0.8\textwidth]{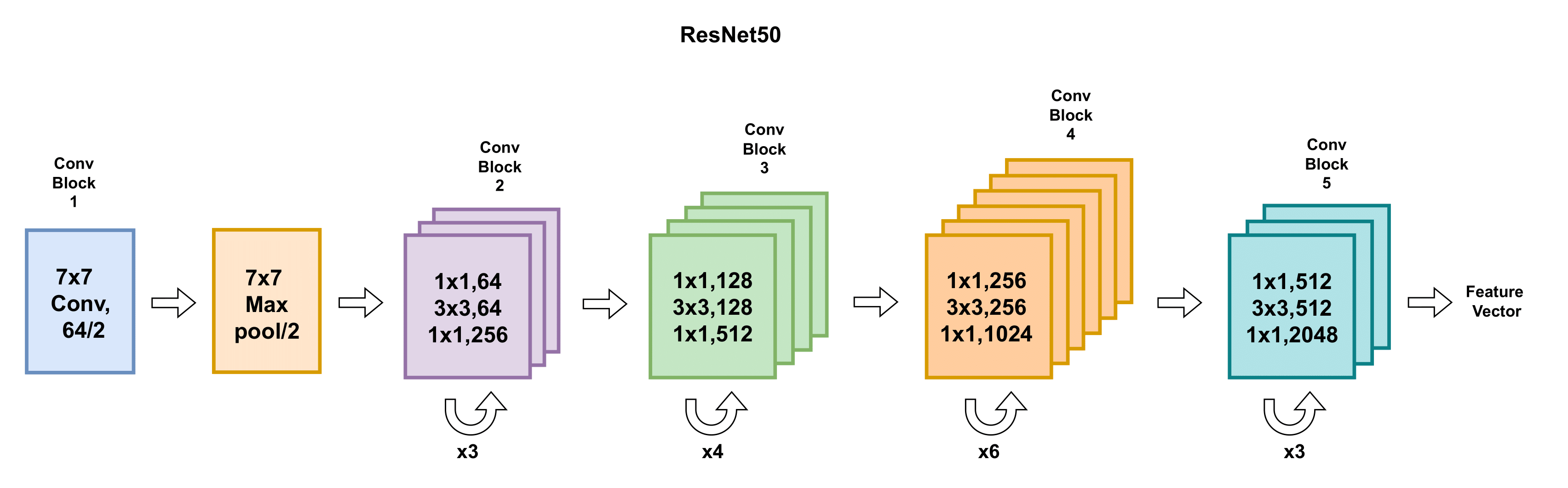}
\caption{Architecture of ResNet50 model}
\label{fig:R_50}
\end{figure*}

\begin{figure*}[!h] 
\centering
\includegraphics[width=0.8\textwidth]{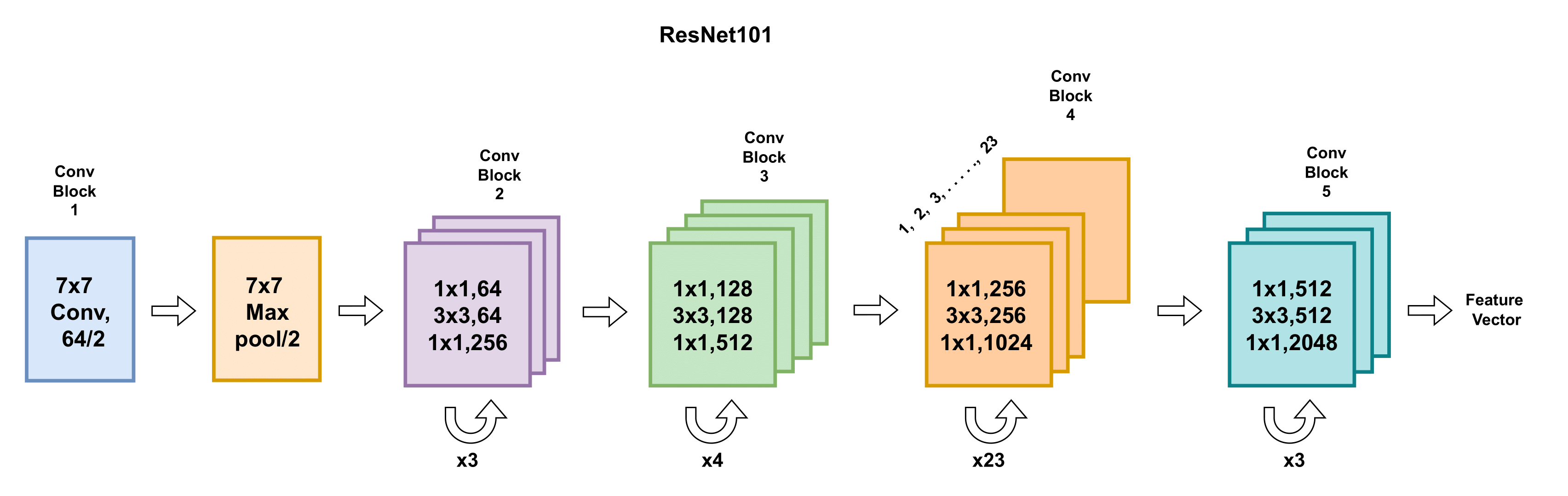}
\caption{Architecture of ResNet101 model}
\label{fig:R_101}
\end{figure*}

\begin{figure*}[!h] 
\centering
\includegraphics[width=0.8\textwidth]{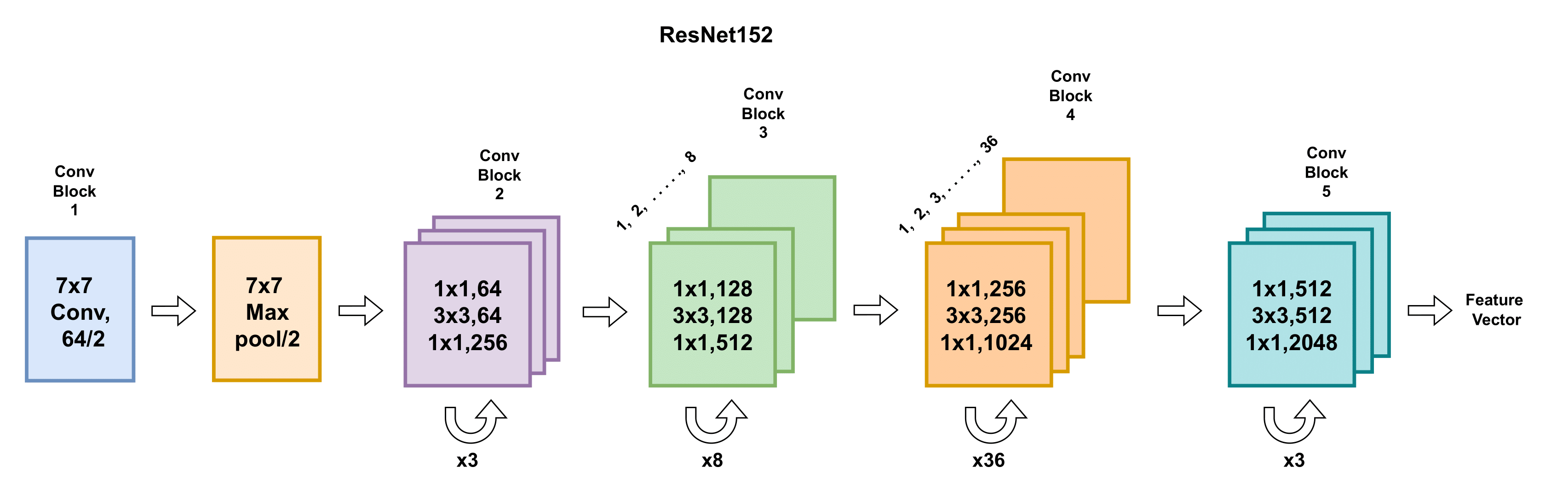}
\caption{Architecture of ResNet152 model}
\label{fig:R_152}
\end{figure*}

ResNets have been recently shown to work well as pretrained deep learning models in the field of medical imaging \cite{hasanah2023deep, toa2024deep, gul2023application}. These models also solve the vanishing gradient problem common in deep neural networks. ResNet50 contains 50 layers in total, which includes 49 convolutional layers that are arranged in 16 residual blocks, as shown in Fig. \ref{fig:R_50}. It is a relatively lightweight model that makes it faster, and it captures simpler patterns. ResNet101 has 101 layers, which include 100 convolutional layers that are organized into 33 residual blocks, as depicted in Fig. \ref{fig:R_101}. It balances depth and complexity, which makes it suitable for tasks where more advanced patterns are to be identified. ResNet152 is the deepest of the three, with 152 layers, including 151 convolutional layers divided into 50 residual blocks, as illustrated in Fig. \ref{fig:R_152}. It is capable of capturing the most complex patterns.

In all three ResNet architectures, we freeze the weights of all convolutional blocks to retain the pretrained knowledge acquired from large-scale datasets such as ImageNet. Each colposcopy image is passed through the network without fine-tuning the original architecture. The final classification layer is removed, and a fixed-length feature vector of size $2048$ is extracted. This feature vector captures high-level semantic information from the input image and is subsequently used for further processing and classification.

Next, we discuss our proposed model, which includes three novel components. \textit{First}, we separately extract detailed and abstract features, which is being explored for the first-time in cancer classification. \textit{Second}, we uniquely embed a non-parametric 3D attention module to enhance these features. \textit{Third}, we propose an innovative adaptive pooling strategy, where GMP is applied to detailed features and GAP to abstract features. These three components are discussed in the following sections.

\paragraph*{Detailed and Abstract Features}
It has been shown in pretrained neural networks that detailed features like edges, color, and texture are captured by earlier convolution blocks, while abstract features like shape and object, are captured by later convolution blocks \cite{naosekpam2022ifvsnet, liu2019novel, tang2017g}. Subsequently, these abstract features dominate the output of the network. Since detailed features of colposcopy cancer image classification are as important as abstract features, extracting features at the end of each convolution block becomes necessary. Hence, for each ResNet, we extract all such features while cascading through the network. In general, ResNets have five convolution blocks. The first two blocks capture the detailed features, and the last three blocks capture the abstract features.

We now explain our feature extraction process. Let the input preprocessed colposcopy image be denoted as \( I \in \mathbb{R}^{C \times H \times W} \), where \( C \), \( H \), and \( W \) represents the number of channels, height, and width, respectively. After passing the input image through the convolution blocks of the ResNet models, we obtain the feature maps as follows:

\begin{equation}
\begin{aligned}
	F_i^{N}\left(c = \{1,...,C\},h = \{1,...,H\},w = \{1,...,W\}\right) &= CB_i^N(I), \\
	\forall~i &\in \{1, \dots, 5\},\\
	\forall~N &\in \{50, 101, 152\}.\\
\end{aligned}
\end{equation}
Here, \( F_i^{N}(c,h,w) \) represents the feature map at the spatial dimension $(h,w)$ and channel position $c$ in the $i^{th}$ convolution block $CB$ associated with the ResNet-$N$ model. 

The channel and the spatial dimensions for the different $CB$ are given as follows:
\begin{align*}
F_1^{N}(c,h,w) &\in \mathbb{R}^{C_1 \times H_1 \times W_1}, \\
&\text{where } C_1 = 64,~ H_1 = 112,~ W_1 = 112, \\
F_2^{N}(c,h,w) &\in \mathbb{R}^{C_2 \times H_2 \times W_2}, \\
&\text{where } C_2 = 256,~ H_2 = 56,~ W_2 = 56, \\
F_3^{N}(c,h,w) &\in \mathbb{R}^{C_3 \times H_3 \times W_3}, \\
&\text{where } C_3 = 512,~ H_3 = 28,~ W_3 = 28, \\
F_4^{N}(c,h,w) &\in \mathbb{R}^{C_4 \times H_4 \times W_4}, \\
&\text{where } C_4 = 1024,~ H_4 = 14,~ W_4 = 14, \\
F_5^{N}(c,h,w) &\in \mathbb{R}^{C_5 \times H_5 \times W_5}, \\
&\text{where } C_5 = 2048,~ H_5 = 7,~ W_5 = 7,\\
&	\forall~N \in \{50, 101, 152\}.
\end{align*}

Further, these feature maps subsequently pass through a non-parametric 3D attention module, which is Euclidean-Pearson Attention Module, to enhance them. We now explain the detailed working of attention module.

\paragraph*{Non-parametric 3D Attention Module} 

The essence of the attention mechanism is to guide the network to concentrate on the most informative regions. Several attention modules are available such as Squeeze-and-Excitation (SE), Convolutional Block Attention Module (CBAM), Simple Attention Module (SimAM), and Euclidean-Pearson Attention Module (EuPea)\cite{CHEN2023106686, pmlr-v139-yang21o, guan2024parameter}. Broadly, attention mechanisms can be categorized into parametric and non-parametric modules. Parametric modules, such as SE and CBAM, rely on learnable parameters (weights) to model feature importance, while non-parametric modules, such as SimAM and EuPea, compute attention directly from the input without learnable parameters. As our feature extraction process does not involve learnable parameters, hence, non-parametric attention modules are a more suitable choice for our framework. Among non-parametric modules, SimAM introduces a 3D attention mechanism that models pixel-wise importance. However, it treats all pixels as equally connected, and ignores the spatial distance between them. This limitation can reduce its effectiveness in capturing fine-grained contextual dependencies.

This limitation becomes especially critical in colposcopy image classification, where visual challenges are more severe than in typical object recognition tasks. Colposcopy images often contain overlapping anatomical structures, reflective glare, and uneven lighting that obscure diagnostically important regions. Additionally, lesions can also vary widely in position, shape, and orientation, making them harder to detect than objects in standard images. Hence, EuPea is a better choice here because it captures spatial relationship between the pixels. Moreover, it also captures the statistical relationship between them that helps in better feature selection. It uses Euclidean distance to consider how far apart pixels are, and a Pearson-inspired correlation to evaluate how different they are in value. Hence, we uniquely integrate this attention module within each convolution block of our model. 

Next, we describe how we calculate attention for pixel value of the colposcopy image using EuPea. Here, for each pixel, the spatial dimension $(h,w)$, channel $c$, convolution block $i$, and ResNet-$N$ model are varied as follows: $h \in \{1, \dots, H\}$, $w \in \{1, \dots, W\}$, $c \in \{1, \dots, C\}$, $i \in \{1, \dots, 5\}$, and $N \in \{50, 101, 152\}$.

First we compute the spatial relationship between pixels as follows: 
\begin{equation}\label{dist}
\begin{aligned}
	D_i^N(c,h,w) &= \sqrt{ \left(F_i^N(c,h,w) - \mu_i^N(c) \right)^2 }, \\
\end{aligned}
\end{equation}
where $\mu_{i}^{N}(c)$ is the mean of all pixels for channel $c$ in the $i^{th}$ convolution block associated with the ResNet-$N$ model. It is calculated as follows:
\begin{equation}
\begin{aligned}
	\mu_{i}^{N}(c) &= \frac{1}{H \times W} \sum_{h=1}^{H} \sum_{w=1}^{W} F_i^{N}(c,h,w). \\
\end{aligned}
\end{equation}

Second, we compute the statistical relationship between the pixels using the Pearson-inspired correlation. It is given as follows:

\begin{equation}\label{cor}
\rho_{i}^{N}(c,h,w) = \frac{(F_i^{N}(c,h,w) - 	\mu_{i}^{N}(c))^2} {4(\left(\sigma_{i}^{N}(c)\right)^2 - \lambda )}+0.5,
\end{equation}
where $\left(\sigma_{i}^{N}(c)\right)^2$ is the variance of all pixels for channel $c$ in the $i^{th}$ convolution block associated with the ResNet-$N$ model. It is calculated as follows:
\begin{equation}
\begin{aligned}
	\left(\sigma_{i}^{N}(c)\right)^2 &= \frac{1}{H \times W} \sum_{h=1}^{H} \sum_{w=1}^{W} \left(F_i^{N}(c,h,w) - \mu_{i}^{N}(c)\right)^2.
\end{aligned}
\end{equation}

The final attention is now defined using a linear combination of the normalized spatial relationship and the statistical relationship between pixels.  This is given as follows:

\begin{equation}
\alpha_{i}^{N}(c,h,w) = \tilde{D}_i^N(c,h,w) + \tilde{\rho}_{i}^{N}(c,h,w)
\end{equation}
where $\tilde{D}_i^N(c,h,w)$ and $\tilde{\rho}_{i}^{N}(c,h,w)$ are the normalized spatial  and statistical relationship, respectively.

Finally, the enhanced features map $\hat{F}_i^{N}(c,h,w)$ is obtained by performing element-wise multiplication between the original feature map and the corresponding attention map (obtained after applying sigmoid on $	\alpha_{i}^{N}(c,h,w)$), as given below,
\begin{equation}\label{dot}
\hat{F}_i^{N}(c,h,w) = sigmoid(\alpha_{i}^{N}(c,h,w)) \odot F_i^{N}(c,h,w).
\end{equation}

\begin{figure*}[!h] 
\centering
\includegraphics[width=0.85\textwidth]{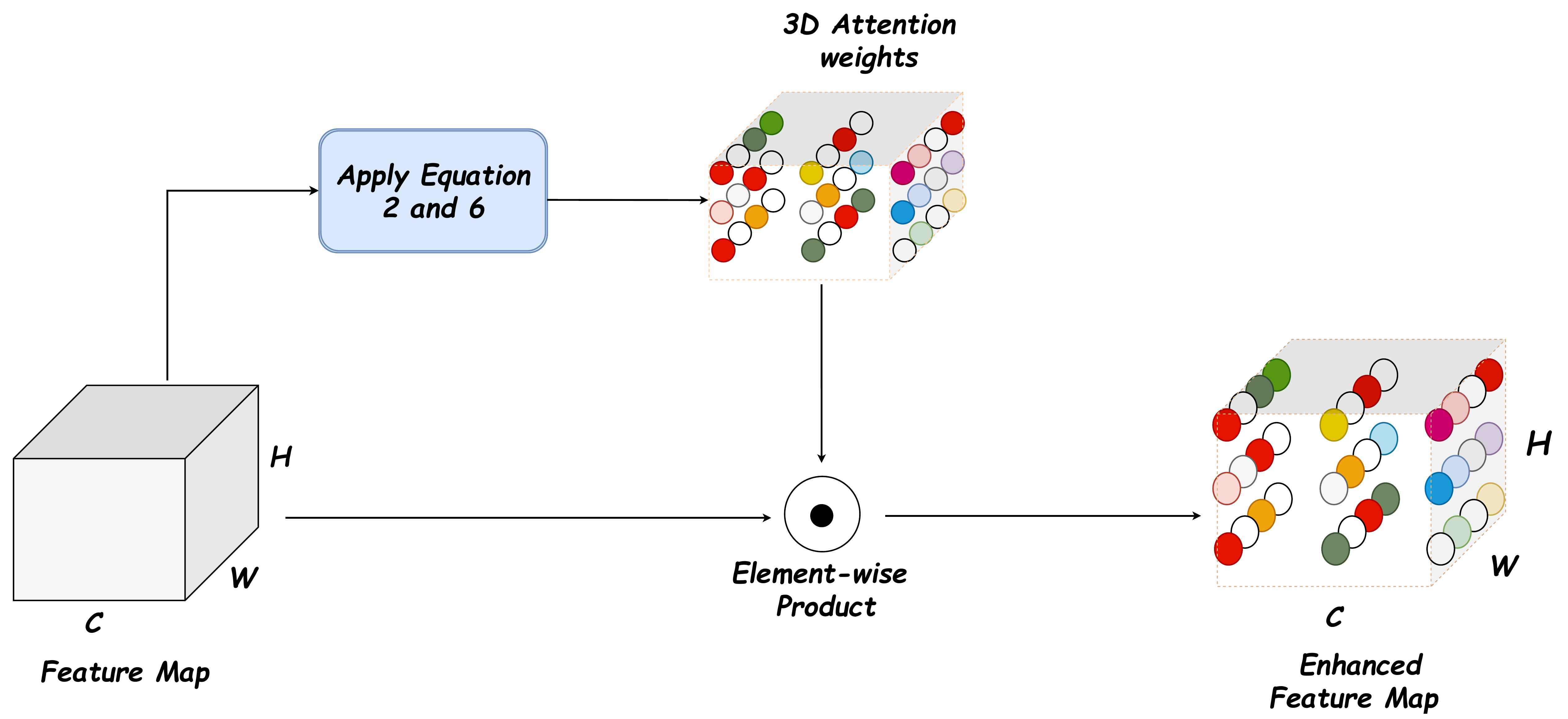}
\caption{Process of applying Non-parametric 3D attention module on feature map}
\label{fig:EuPea}
\end{figure*}

Fig. \ref{fig:EuPea} depicts the steps of applying attention on a feature map. Next, the enhanced feature maps are passed through our proposed adaptive pooling component, which consists of GMP and GAP layer, as discussed in the following section.

\paragraph*{Adaptive Pooling} 
In this section, we select the features from the enhanced feature maps. This can be performed in many ways. In deep learning, one of the common approaches is pooling, which can be done using various methods such as Global Sum Pooling (GSP), Global Max Pooling (GMP), and Global Average Pooling (GAP) \cite{aich2018global, tao2022pooling}. As discussed earlier, GMP is effective in retaining the strongest activations by filtering out the noise. The initial two convolution blocks that provide detailed features also carry background noise. Hence, to eliminate this noise, we pass the output of the first two convolution blocks through GMP layers. 

The GAP layer averages out the features over the whole image and does not filter out any background noise. The later three convolution blocks that provide abstract features do not have any noise from the background. Hence, we pass the output of the last three convolution blocks through GAP layers \cite{zhao2024improved, nirthika2022pooling}. This novel adaptive pooling strategy improves both feature relevance and model efficiency and is formulated as follows: 

\begin{equation}
f_i^{N} = 
\begin{cases}
	\text{GMP}(\hat{F}_i^{N}(c,h,w)) = \max\limits_{h, w} \hat{F}_i^{N}(c, h, w), \\
	\hspace{8em} \forall ~i = \{1, 2\}, \\
	\text{GAP}(\hat{F}_i^{N}(c,h,w)) = \dfrac{1}{H \cdot W} \sum\limits_{h=1}^{H} \sum\limits_{w=1}^{W} \hat{F}_i^{N}(c, h, w), \\
	\hspace{8em} \forall ~i = \{3, 4, 5\},
\end{cases}
\end{equation}

where the range of $c$, $h$, and $w$ are same as defined earlier, $f_i^{N}$ is the pooled feature vector from the $i^{th}$ convolution block of the ResNet-\(N\) model.
The final feature vector is formed by fusing the GMP and GAP feature vectors obtained from each convolution block and represented as follows:
\begin{equation}
f^{N} =  f_1^{N} \, \| \, f_2^{N} \, \| \, f_3^{N} \, \| \, f_4^{N} \, \| \, f_5^{N} ,
\end{equation}
where $\|$ denotes vector concatenation operation. 

\begin{figure*}[!htbp] 
\centering
\includegraphics[width=\textwidth]{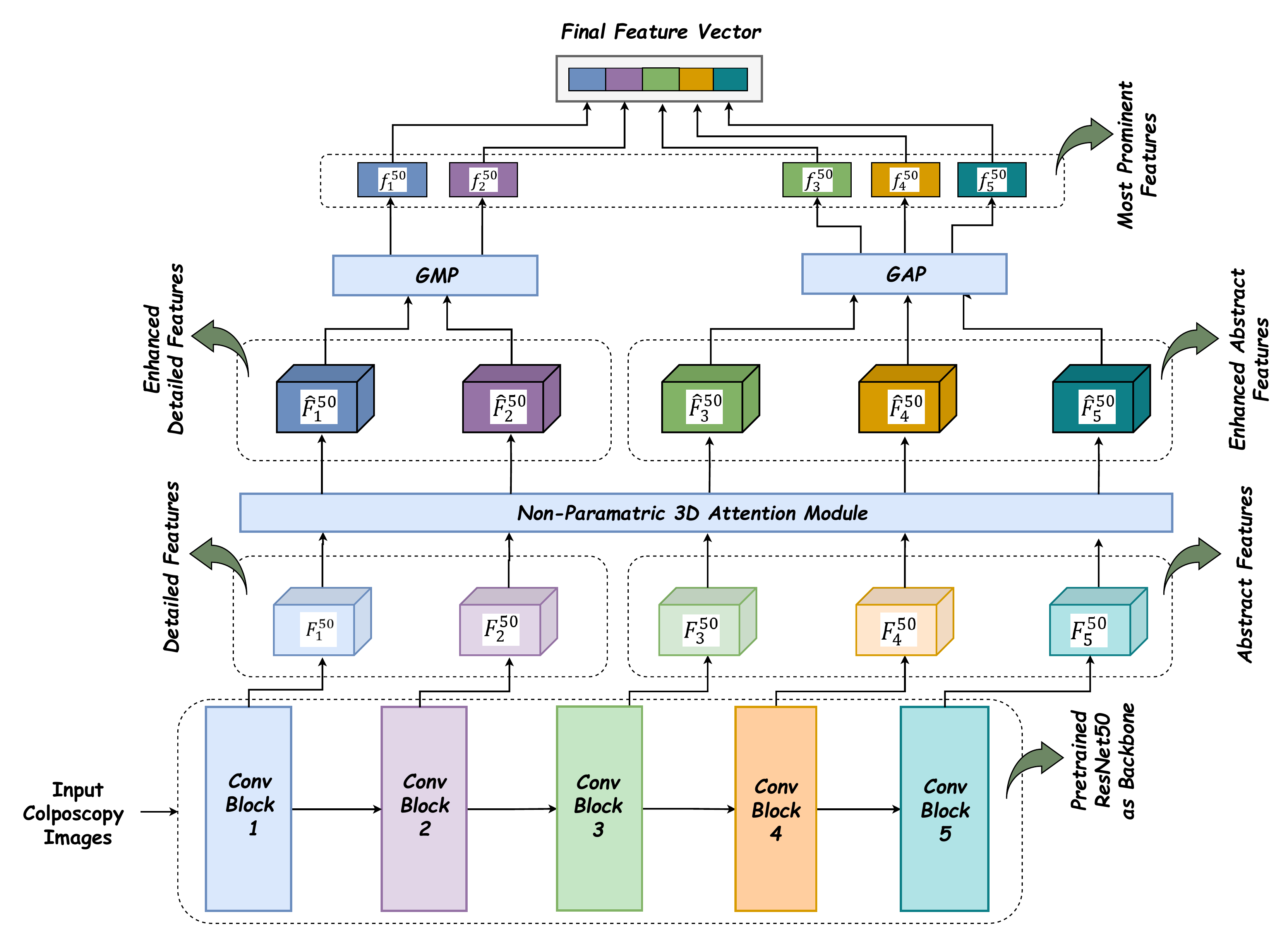}
\caption{Block-Fused Attention-Driven Adaptively-Pooled ResNet (BF-AD-AP-ResNet) model}
\label{fig:BF-AD-AP-ResNet50}
\end{figure*}

These three components (extraction of detailed and abstract features, a non-parametric 3D attention module, and adaptive pooling) work together to form our novel BF-AD-AP-ResNets models, which include BF-AD-AP-ResNet50, BF-AD-AP-ResNet101, and BF-AD-AP-ResNet152, corresponding to ResNet50, ResNet101, and ResNet152, respectively. Fig. \ref{fig:BF-AD-AP-ResNet50} illustrates the architecture of our proposed  BF-AD-AP-ResNet50 model. The architectures for  BF-AD-AP-ResNet101 and  BF-AD-AP-ResNet152 models can be derived by substituting the backbone network accordingly. 

\subsubsection{Tri-Stream version of our proposed models}

In this section, we discuss the working of our proposed and more powerful Tri-Stream model, which consolidates the features from our three proposed models (BF-AD-AP-ResNet50, BF-AD-AP-ResNet101, and BF-AD-AP-ResNet152). Next we present the mathematical formulation of Proposed Tri-Stream model.

Let $f^{50}$, $f^{101}$, and $f^{152}$ denote the feature vectors extracted from the BF-AD-AP-ResNet50, BF-AD-AP-ResNet101, and BF-AD-AP-ResNet152 models, respectively. These vectors are concatenated to form our Proposed Tri-Stream model, which can be mathematically represented as follows:

\begin{equation}
\mathbf{f_{tri-stream}} = f^{50} \, \| \, f^{101} \, \| \, f^{152},
\end{equation}
where $\|$ denotes the concatenation operation. 

This approach enables the model to benefit from diverse feature representation, which helps to improve classification performance across diverse colposcopy image variations.

The complete step-by-step process, including feature extraction from each convolution block, applying attention to these features, selection using adaptive pooling, fusion of features across all blocks, and final concatenation of features from multiple BF-AD-AP-ResNets to form the Proposed Tri-Stream model, is outlined in Algorithm \ref{Algo}.
\begin{algorithm}[!htbp]
\caption{Proposed BF-AD-AP-ResNets based Tri-Stream Model}
\label{Algo}
\begin{algorithmic}[1]
	\Require Input preprocessed colposcopy image $I \in \mathbb{R}^{C \times H \times W}$
	\Ensure Final concatenated feature vector $\mathbf{f_{(tri-stream)}}$
	
	\State \textbf{Initialize:} Pretrained ResNet models with depths $N \in \{50, 101, 152\}$
	
	\For{each ResNet-$N$ model}
	\For{$i = 1$ to $5$ (convolutional blocks)}
	\State Extract feature map $F_i^{N}(c,h,w) = \text{CB}_i(I)$
	\State Apply EuPea attention to $F_i^{N}(c,h,w)$ using Eq. \ref{dist} to \ref{dot} to get $\hat{F}_i^{N}(c,h,w)$
	\If{$i \in \{1, 2\}$}
	\State $f_i^{N} = \text{GMP}(\hat{F}_i^{N}(c,h,w))$ \Comment{Global Max Pooling}
	\Else
	\State $f_i^{N} = \text{GAP}(\hat{F}_i^{N}(c,h,w))$ \Comment{Global Average Pooling}
	\EndIf
	\EndFor
	\State Concatenate block-level vectors: 
	\[
	f^{N} = f_1^{N} \, \| \, f_2^{N} \, \| \, f_3^{N} \, \| \, f_4^{N} \, \| \, f_5^{N}
	\]
	\EndFor
	
	\State Concatenate features from all three BF-AD-AP-ResNets models: 
	\[
	\mathbf{f_{tri-stream}} = f^{50} \, \| \, f^{101} \, \| \, f^{152}
	\]
	\State \Return $\mathbf{f_{tri-stream}}$
\end{algorithmic}
\end{algorithm}

Next, the data is split into training and testing parts (coming from cross-validation, which is discussed in the Results section). Subsequently, features are normalized, as discussed below.  

\subsection{Feature normalization}
We normalize the features to prevent the classification model from biased learning and ensure fair comparisons between different units or measurement scales. There are various ways available to normalize the features, i.e., z-scored normalization, variable stability scaling, min-max normalization, etc. \cite{singh2020investigating}. For us, the min-max normalization technique, which scales the features in the range between 0 and 1, works the best. It is formulated as follows:
\begin{equation}
s_{\text{scaled}} = \frac{s - s_{\text{min}}}{s_{\text{max}} - s_{\text{min}}} ,
\end{equation}
where $s_{\text{scaled}}$ is the scaled value of a feature, $s$ is the original value of the feature, $s_{\text{min}}$ is the minimum, and $s_{\text{max}}$ is the maximum value among all the feature values. Next, the data is fed to the classifier, which is discussed below.

\subsection{Classification}

Numerous machine learning classifiers are available for classification tasks, including K-Nearest Neighbors (KNN), Logistic Regression (LR), Naïve Bayes (NB), Neural Network (NN), and Support Vector Machine (SVM) \cite{saidi2025efficient}. In this work, we choose SVM as our classification algorithm because it works well with high-dimensional data \cite{cortes1995support}, which is important since our proposed BF-AD-AP-ResNets produce a large and combined feature vector, $\mathbf{f_{tri-stream}}$. Moreover, it is known for its robustness to overfitting, especially in cases where the number of features significantly exceeds the number of samples. 

SVM is a supervised learning method used to classify data. It works by finding a function to capture the relationships between different points, which is eventually used to separate them into classes. As mentioned above, for the types and the CIN score classification, we have three classes. The standard SVM performs a two-class classification. This can be easily generalized to a three-class classification by using a combination of classes.

\section{Experimental results}\label{sec: Result}

In this section, we provide the numerical results of our model on two datasets. For training and testing, we use 5-fold and 10-fold cross-validation methods. In 5-fold, the data is split into roughly five equal parts, known as folds. Now, the model is trained and tested five times, with a different fold used as the test set in each round, while the other four folds are used for training. After all five rounds, the results are averaged to give a more reliable and unbiased measure of the model performance on the new data. Similarly, in 10-fold cross-validation, the process is the same, but the data is split into ten folds, and the model is trained and tested across ten iterations. To assess the performance of our system, we use standard metrics such as sensitivity, specificity, precision, F1-score, and accuracy. 

Sensitivity ($Sens$) quantifies the ability of the model to correctly identify positive instances, while Specificity ($Spec$) measures the ability to correctly detect negative instances. They are computed as follows:

\begin{equation}
Sens = \frac{TP}{TP + FN}; \quad Spec = \frac{TN}{TN + FP},
\end{equation}
where TP, TN, FP, and FN mean True Positive, True Negative, False Positive, and False Negative, respectively.

Precision ($Pre$) refers to the proportion of correctly predicted positive instances out of all instances that were predicted as positive while F1-score ($F1$) computes harmonic mean of sensitivity and precision, offering a balanced measure when both false positives and false negatives are important. Mathematically, they are calculated as follows:\\

\begin{equation}
Pre = \frac{TP}{TP + FP}; \quad F1 = \frac{2 \times Pre \times Sens}{Pre + Sens}.
\end{equation}


Accuracy ($Acc$) represents the overall correctness of the classification model and is calculated as the ratio of the correctly predicted instances to the total instances:
\begin{equation}
Acc = \frac{TP + TN}{TP + FN +  TN + FP }.
\end{equation}

\begin{figure*}[!htbp]
\centering
\begin{subfigure}[b]{0.4\textwidth}
	\includegraphics[width=\textwidth]{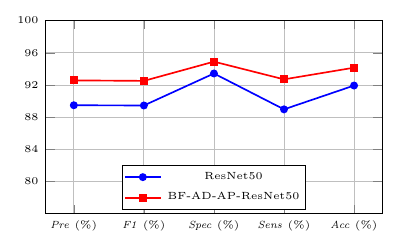}
	\caption{}
	\label{fig:5FR50}
\end{subfigure}
\begin{subfigure}[b]{0.4\textwidth}
	\includegraphics[width=\textwidth]{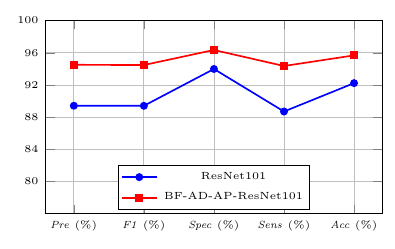}
	\caption{}
	\label{fig:5FR101}
\end{subfigure}
\begin{subfigure}[b]{0.4\textwidth}
	\includegraphics[width=\textwidth]{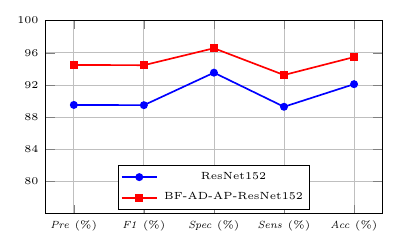}
	\caption{}
	\label{fig:5FR152}
\end{subfigure}
\caption{Performance line plot for the types classification with 5-fold cross-validation on IARC dataset. (a), (b), and (c) depicts the one to one comparison of Base ResNets model with Proposed BF-AD-AP-ResNets models}
\label{fig:bar_types_5f}
\end{figure*}

\begin{figure*}[!htbp]
\centering
\begin{subfigure}[b]{0.4\textwidth}
	\includegraphics[width=\textwidth]{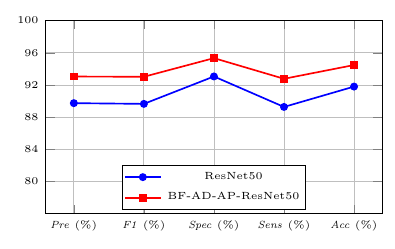}
	\caption{}
	\label{fig:10FR50}
\end{subfigure}
\begin{subfigure}[b]{0.4\textwidth}
	\includegraphics[width=\textwidth]{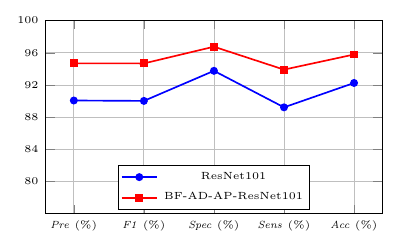}
	\caption{}
	\label{fig:10FR101}
\end{subfigure}
\begin{subfigure}[b]{0.4\textwidth}
	\includegraphics[width=\textwidth]{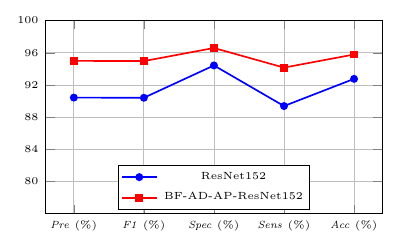}
	\caption{}
	\label{fig:10FR152}
\end{subfigure}

\caption{Performance line plot for the types classification with 10-fold cross-validation on IARC dataset. (a), (b), and (c) depicts the one to one comparison of Base ResNets with Proposed BF-AD-AP-ResNets models}
\label{fig:bar_types_10f}
\end{figure*}

The results for the types classification using 5-fold and 10-fold cross-validation on the IARC dataset are presented in Table \ref{tab:result_type_5f} and Table \ref{tab:result_type_10f}, respectively. Here, the first column lists the feature extraction techniques, and the rest of the columns list the values of the performance metrics. The corresponding line plots are given in Fig. \ref{fig:bar_types_5f} and Fig. \ref{fig:bar_types_10f}, respectively.
 
\renewcommand{\arraystretch}{1.4}
\begin{table*}[!htbp]

\centering
\scriptsize
\caption{Quantitative performance comparison of the proposed models using 5-fold cross-validation for the types classification on IARC dataset}
\vspace{5pt}
\resizebox{\textwidth}{!}{
	\begin{tabular}{l|c|c|c|c|c||c}
		\hline
		Features Extraction Techniques   &$Pre(\%)$  & $F1(\%) $& $Spec (\%) $  &$ Sens (\%) $  & $Acc (\%) $ & $\textit{\textbf{Avg (\%)}}$\\ \hline

		ResNet50                      		& 89.48				& 89.45				& 93.43            & 88.97          	 & 91.93   & 90.65   \\ 
		\textbf{BF-AD-AP-ResNet50}                       & \textbf{92.57}    & \textbf{92.51}    & \textbf{94.90}   & \textbf{92.70}      & \textbf{94.15} & \textbf{93.37}\\ \hline
		
		ResNet101  							& 89.42				& 89.41				& 93.99            & 88.70               & 92.23 &90.75 \\ 
		\textbf{BF-AD-AP-ResNet101}                      & \textbf{94.51}    & \textbf{94.48}    & \textbf{96.35}   & \textbf{94.36}      & \textbf{95.68} &\textbf{95.08}\\ \hline

		ResNet152                           & 89.52             & 89.49             & 93.53            & 89.29               & 92.10 & 90.79       \\ 
		\textbf{BF-AD-AP-ResNet152}                     & \textbf{94.49}    & \textbf{94.46}    & \textbf{96.57}   & \textbf{93.24}      & \textbf{95.47} & \textbf{94.85}\\ \hline
							
		\textbf{Proposed Tri-Stream Model }  	& \textbf{98.57}	& \textbf{98.58}	& \textbf{98.86}   & \textbf{98.47}      & \textbf{98.73} & \textbf{98.64} \\ \hline

	\end{tabular}
	
	\label{tab:result_type_5f}
}
\end{table*}

\renewcommand{\arraystretch}{1.4}
\begin{table*}[!htbp]

\centering
\scriptsize
\caption{Quantitative performance comparison of the proposed models using 10-fold cross-validation for the types classification on IARC dataset}
\vspace{5pt}
\resizebox{\textwidth}{!}{
	\begin{tabular}{l|c|c|c|c|c||c}
		\hline
		Features Extraction Techniques    &$Pre(\%)$  & $F1(\%) $& $Spec (\%) $  &$ Sens (\%) $  & $Acc (\%) $ & $\textit{\textbf{Avg (\%)}}$ \\ \hline

		ResNet50  						  &  89.74             & 89.65             & 93.06           & 89.27            & 91.80  & 90.70   \\ 
		\textbf{BF-AD-AP-ResNet50}					  &  \textbf{93.05}    & \textbf{ 93.02}   & \textbf{95.34}  & \textbf{92.77}   & \textbf{94.48} & \textbf{93.73}\\ \hline
		
		ResNet101  						  &  90.07 			   & 90.02 			   & 93.76           & 89.22            & 92.25  &  91.06 \\ 
		\textbf{BF-AD-AP-ResNet101}					  & \textbf{94.69}	   & \textbf{94.68}    & \textbf{96.76}  & \textbf{93.91}   & \textbf{95.79} & \textbf{95.17}\\ \hline

		ResNet152     				 	  &  90.43  		   & 90.41             & 94.43           & 89.38            & 92.76     & 91.48   \\ 
		\textbf{BF-AD-AP-ResNet152}					  &\textbf{95.00} 	   & \textbf{94.97}    & \textbf{96.60}  & \textbf{94.16}   & \textbf{95.79}& \textbf{95.30}\\ \hline

		\textbf{Proposed Tri-Stream Model}  &\textbf{98.50}      & \textbf{98.51 }   & \textbf{98.75}  & \textbf{98.63}   & \textbf{98.71}    & \textbf{98.62}    \\  
		\hline

\end{tabular}}

\label{tab:result_type_10f}

\end{table*}

It is apparent from the tables that Base ResNets, Proposed BF-AD-AP-ResNets, and Proposed Tri-Stream model achieves an average performance of $90.91\%$, $94.58\%$, and $\textbf{98.63\%}$, respectively. Thus we can see we achieve  $\approx\textbf{8\%}$ gain. We compute this average performance by taking the mean of all five performance metrics obtained from both 5-fold and 10-fold cross-validation. 

We also compare our results with the 10-fold cross-validation results reported in \cite{dash2023cervical}. Our method shows a substantial improvement over theirs, with gains of $\textbf{10\%}$ in precision, $\textbf{17\%}$ in F1-score, $\textbf{8\%}$ in specificity, $\textbf{17\%}$ in sensitivity, and $\textbf{17\%}$ in accuracy.

\begin{figure*}[!htbp]
\centering
\begin{subfigure}[b]{0.4\textwidth}
	\includegraphics[width=\textwidth]{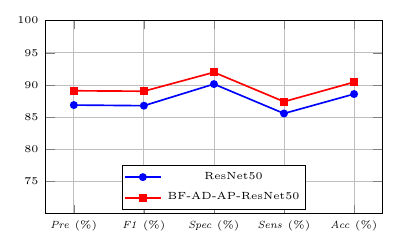}
	\caption{}
	\label{fig:5FR50N}
\end{subfigure}
\begin{subfigure}[b]{0.4\textwidth}
	\includegraphics[width=\textwidth]{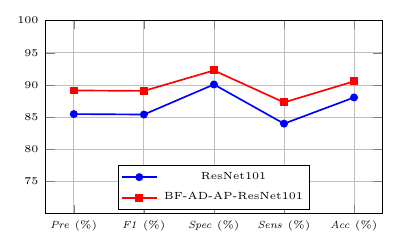}
	\caption{}
	\label{fig:5FR101N}
\end{subfigure}
\begin{subfigure}[b]{0.4\textwidth}
	\includegraphics[width=\textwidth]{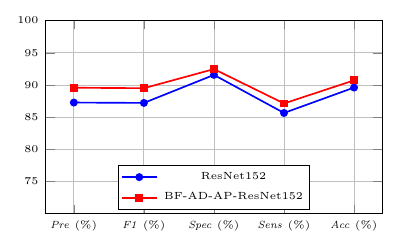}
	\caption{}
	\label{fig:5FR152N}
\end{subfigure}
\caption{Performance line plot for the CIN score classification with 5-fold cross-validation on AnnoCerv dataset. (a), (b), and (c) depicts the one to one comparison of Base ResNets with Proposed BF-AD-AP-ResNets models}
\label{fig:bar_cin_5f}
\end{figure*}

\begin{figure*}[!htbp]
\centering
\begin{subfigure}[b]{0.4\textwidth}
	\includegraphics[width=\textwidth]{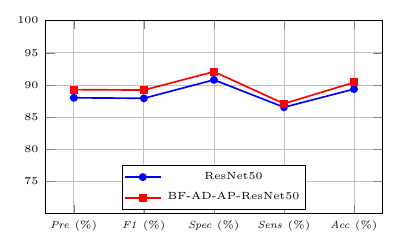}
	\caption{}
	\label{fig:10FR50N}
\end{subfigure}
\begin{subfigure}[b]{0.4\textwidth}
	\includegraphics[width=\textwidth]{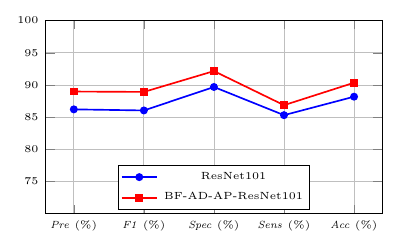}
	\caption{}
	\label{fig:10FR101N}
\end{subfigure}
\begin{subfigure}[b]{0.4\textwidth}
	\includegraphics[width=\textwidth]{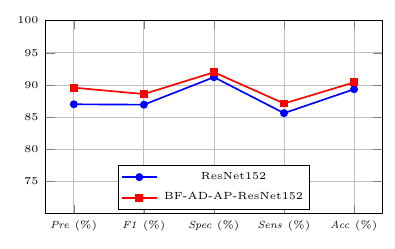}
	\caption{}
	\label{fig:10FR152N}
\end{subfigure}

\caption{Performance line plot for the CIN score classification with 10-fold cross-validation on AnnoCerv dataset. (a), (b), and (c) depicts the one to one comparison of Base ResNets with Proposed BF-AD-AP-ResNets models}
\label{fig:bar_cin_10f}
\end{figure*}

Next, the results for the CIN score classification on the AnnoCerv dataset using 5-fold and 10-fold cross-validation are given in Table \ref{tab:result_CIN_5f} and Table \ref{tab:result_CIN_10f}, respectively. The respective line plots are given in Fig. \ref{fig:bar_cin_5f} and Fig. \ref{fig:bar_cin_10f}. This follows the same pattern as the types classification. 

\renewcommand{\arraystretch}{1.4}
\begin{table*}[!htbp]
\scriptsize
\centering

\caption{Quantitative performance comparison of the proposed models for the CIN score classification with 5-fold cross-validation on AnnoCerv dataset}
\vspace{5pt}
\resizebox{\textwidth}{!}{
	\begin{tabular}{l|c|c|c|c|c||c}
		\hline
		Features Extraction Techniques &$Pre(\%)$  & $F1(\%) $& $Spec (\%) $  &$ Sens (\%) $  & $Acc (\%) $ & $\textit{\textbf{Avg (\%)}}$ \\ 
		\hline
		
		ResNet50                              & 86.88             &  86.78              & 90.14                & 85.57               & 88.59  & 87.59    \\ 
		\textbf{BF-AD-AP-ResNet50}                         & \textbf{89.10}    & \textbf{89.03}    & \textbf{91.97}    & \textbf{87.41}   & \textbf{90.44}   & \textbf{89.59}  \\ \hline
		
		ResNet101                            & 85.48             & 85.41              & 90.08              & 83.99             & 88.06 &    86.60  \\ 
		\textbf{BF-AD-AP-ResNet101}                      & \textbf{89.15}    &  \textbf{89.08}     & \textbf{92.26}    & \textbf{87.29}    & \textbf{90.57}    & \textbf{89.67}  \\ \hline
		
		ResNet152                           & 87.27             &  87.21            & 91.57                & 85.65            & 89.60   & 88.26     \\ 
		\textbf{BF-AD-AP-ResNet152}                       &\textbf{89.57}        &  \textbf{89.50}   & \textbf{92.47}  & \textbf{87.13}    &\textbf{90.72}     & \textbf{89.88}   \\ \hline

		\textbf{Proposed Tri-Stream Model}     &\textbf{93.71}       & \textbf{93.63}      & \textbf{94.51}      &\textbf{ 92.90}   & \textbf{93.95}    &\textbf{93.74}    \\
		\hline

\end{tabular}}

\label{tab:result_CIN_5f}

\end{table*}

\renewcommand{\arraystretch}{1.2}
\begin{table*}[!htbp]

\centering

\caption{Quantitative performance comparison of the proposed models for the CIN score classification with 10-fold cross-validation on AnnoCerv dataset}
\vspace{5pt}
\resizebox{\textwidth}{!}{
	\begin{tabular}{l|c|c|c|c|c||c}
		\hline
		Features Extraction Techniques &$Pre(\%)$  & $F1(\%) $& $Spec (\%) $  &$ Sens (\%) $  & $Acc (\%) $ & $\textit{\textbf{Avg (\%)}}$\\ 
		\hline
		
		ResNet50                              & 88.01             &  87.93              & 90.80               & 86.54               & 89.36  & 88.53     \\ 
		\textbf{BF-AD-AP-ResNet50}                         & \textbf{89.28}    & \textbf{89.21}    & \textbf{92.06}    & \textbf{87.10}   & \textbf{90.40}  &  \textbf{89.61}  \\ \hline
		
		ResNet101                            & 86.21             & 86.04              & 89.69              & 85.31           & 88.18    & 87.09 \\ 
		\textbf{BF-AD-AP-ResNet101}                & \textbf{88.98}    &  \textbf{88.92}     & \textbf{92.16}    & \textbf{86.86}    & \textbf{90.35}   &   \textbf{89.45}    \\ \hline
		
		ResNet152                           & 87.00             &  86.94            & 91.21                & 85.62            & 89.34    & 88.02    \\ 
		\textbf{BF-AD-AP-ResNet152}                      &\textbf{89.58}        &  \textbf{89.59}   & \textbf{91.98}  & \textbf{87.12}    &\textbf{90.40} & \textbf{89.73}       \\ \hline
		
		\textbf{Proposed Tri-Stream Model}     &\textbf{93.04}       & \textbf{92.71}      & \textbf{92.41}      &\textbf{ 94.08}   & \textbf{92.96}  & \textbf{93.04}      \\
		\hline

\end{tabular}}

\label{tab:result_CIN_10f}

\end{table*}

Here, Base ResNets, Proposed BF-AD-AP-ResNets, and Proposed Tri-Stream model achieves an average performance of $87.68\%$, $89.66\%$, and $\textbf{93.39\%}$, respectively. Again, our gain is $\approx\textbf{6\%}$. The average performance is computed in the same manner as described for the IARC dataset. Currently, no competitive work is available on this dataset for comparison.

\section{Ablation Study}\label{sec: ablation}
In this section, we conduct a comprehensive ablation study to evaluate the contribution of each component in the Proposed BF-AD-AP-ResNet model. Here, we use BF-AD-AP-ResNet101 model among the three available ones. Without loss of generality this idea can be applied for the other two (BF-AD-AP-ResNet50 and BF-AD-AP-ResNet152) as well. The evaluation is conducted on both datasets, under both 5-fold and 10-fold cross-validation settings.

Table \ref{tab:ablation_iarc_5f} and Table \ref{tab:ablation_iarc_10f} presents the results of our ablation study on the IARC dataset using 5-fold and 10-fold cross-validation, respectively. Here, the Base ResNet101 model, which utilizes only abstract features, achieves an average performance of $90.75\% $. In the proposed study, when both abstract and detailed features are extracted and selected by GMP layer, the performance improves to $92.65\% $. Replacing GMP with GAP leads to further gains, raising the average performance to $93.06\% $. When we apply our Proposed adaptive pooling, the performance increases to $94.06\%$. Finally, integrating the EuPea attention module with adaptive pooling, which is Proposed BF-AD-AP-ResNet101, enables the model to achieve its best average performance of $95.08\%$. This average is calculated by using the value of average performance of both 5-fold and 10-fold cross validation.

\renewcommand{\arraystretch}{1.3}
\begin{table*}[!htbp]

\centering

\caption{Ablation study results on the IARC dataset using 5-fold cross-validation}
\vspace{5pt}
\resizebox{\textwidth}{!}{
	\begin{tabular}{l|c|c|c|c|c||c}
		\hline
		Features Extraction Techniques &$Pre(\%)$  & $F1(\%) $& $Spec (\%) $  &$ Sens (\%) $  & $Acc (\%) $ & $\textit{\textbf{Avg (\%)}}$ \\ 
		\hline
		
		Base ResNet101      		& 89.42			& 89.41           & 93.99        	& 88.70        	& 92.23    &   \textbf{90.75} \\
		
		Block-Fused ResNet101 (All GMP) & 91.98 			& 91.94            & 94.44        	& 91.47       	& 93.46    & \textbf{92.65 } \\
		Block-Fused ResNet101 (All GAP) & 92.78			&92.75            & 94.85        	& 91.24        	& 93.68   & \textbf{93.06} \\
		Block-Fused ResNet101 (AP)& 93.54 			& 93.49             & 95.45         	& 93.14        	& 94.70   &  \textbf{94.06 }\\
		Block-Fused ResNet101 (EuPea + AP) & 94.51 & 94.48 & 96.35 & 94.36 & 95.68 & \textbf{95.08 }\\
		\hline

\end{tabular}}

\label{tab:ablation_iarc_5f}

\begin{tablenotes}

	\scriptsize
	\item AP stand for Adaptive Pooling.
\end{tablenotes}
\end{table*}

\renewcommand{\arraystretch}{1.3}
\begin{table*}[!htbp]

\centering

\caption{Ablation study results on the IARC dataset using 10-fold cross-validation}
\vspace{5pt}
\resizebox{\textwidth}{!}{
		\begin{tabular}{l|c|c|c|c|c||c}
		\hline
		Features Extraction Techniques &$Pre(\%)$  & $F1(\%) $& $Spec (\%) $  &$ Sens (\%) $  & $Acc (\%) $ & $\textit{\textbf{Avg (\%)}}$\\ 
		\hline
		
		Base ResNet101      		& 90.07		& 90.02          & 93.76        	& 89.22        	& 92.25    &   \textbf{91.06} \\
		
		Block-Fused ResNet101 (All GMP) & 92.19			& 92.14           & 94.85       	& 90.99       	& 93.56   & \textbf{92.75 }\\
		Block-Fused ResNet101 (All GAP) & 92.80			&92.81           & 95.13       	& 92.62       	& 93.60  & \textbf{93.39 }\\
		Block-Fused ResNet101 (AP)& 93.90			& 93.89            & 95.74         	& 93.65       	& 95.02   &  \textbf{94.44 }\\
		Block-Fused ResNet101 (EuPea + AP) & 94.69 & 94.68 & 96.76 & 93.91 & 95.79 & \textbf{95.17} \\
		\hline

		\end{tabular}}

\label{tab:ablation_iarc_10f}

\end{table*}

Tables \ref{tab:ablation_annocerv_5f} and Table \ref{tab:ablation_annocerv_10f} report the results on the AnnoCerv dataset under 5-fold and 10-fold cross-validation, respectively. A similar performance trend is observed, where each successive enhancement leads to consistent improvement. The Base ResNet101 model achieves an average performance of $91.06\%$. In the proposed work, the use of GMP yields a clear gain, increasing performance to $92.75\%$, while replacing GMP with GAP further improves it to $93.39\%$. Applying adaptive pooling raises the results to $94.44\%$. The best performance is achieved when the EuPea attention module is combined with adaptive pooling, reaching an average of $95.17\%$. The averages are computed using the same method as described above. 

\renewcommand{\arraystretch}{1.3}
\begin{table*}[!htbp]

\centering

\caption{Ablation study results on the AnnoCerv dataset using 5-fold cross-validation}
\vspace{5pt}
\resizebox{\textwidth}{!}{
		\begin{tabular}{l|c|c|c|c|c||c}
		\hline
		Features Extraction Techniques &$Pre(\%)$  & $F1(\%) $& $Spec (\%) $  &$ Sens (\%) $  & $Acc (\%) $ &$\textit{\textbf{Avg (\%)}}$ \\ 
		\hline
		
		Base ResNet101      		& 85.48			& 85.41           & 90.08        	& 83.99       	& 88.06    &   \textbf{86.60} \\
		
		Block-Fused ResNet101 (All GMP) & 87.40 			& 87.36          & 91.05       	& 84.25       	& 88.78    & \textbf{87.77}  \\
		Block-Fused ResNet101 (All GAP) & 87.95		&87.91            & 90.94        	& 84.70      	& 88.85   & \textbf{88.07} \\
		Block-Fused ResNet101 (AP)& 88.58 			& 88.48            & 91.47        	& 86.29        	& 89.75  &  \textbf{88.91 }\\
		Block-Fused ResNet101 (EuPea + AP) & 89.15 & 89.08 & 92.26 & 87.29 & 90.57 & \textbf{89.67} \\
		\hline

		\end{tabular}}

\label{tab:ablation_annocerv_5f}

\end{table*}

\renewcommand{\arraystretch}{1.3}
\begin{table*}[!htbp]

\centering

\caption{Ablation study results on the AnnoCerv dataset using 10-fold cross-validation}
\vspace{5pt}
\resizebox{\textwidth}{!}{
		\begin{tabular}{l|c|c|c|c|c||c}
		\hline
		Features Extraction Techniques &$Pre(\%)$  & $F1(\%) $& $Spec (\%) $  &$ Sens (\%) $  & $Acc (\%) $ & $\textit{\textbf{Avg (\%)}}$\\ 
		\hline
		
		Base ResNet101      		& 86.21			& 86.04          & 89.69       	& 85.31       	& 88.18    &   \textbf{87.09 }\\
		
		Block-Fused ResNet101 (All GMP) & 87.65 			& 87.60            & 91.28        	& 85.17       	& 89.26    & \textbf{88.19}  \\
		Block-Fused ResNet101 (All GAP) & 88.63		&88.53           & 91.26       	& 85.77     	& 89.43  & \textbf{88.72}\\
		Block-Fused ResNet101 (AP)& 88.69 			& 88.58           & 90.97        	& 86.82        	& 89.60   &  \textbf{88.93} \\
	Block-Fused ResNet101 (EuPea + AP)&88.98	& 88.92&92.16&86.86  & 90.35    &   \textbf{89.45} \\
		\hline

		\end{tabular}}

\label{tab:ablation_annocerv_10f}

\end{table*}

\section{Explainable AI (XAI)}\label{sec:xai}
XAI refers to the process of understanding and explaining the decisions of machine learning or deep learning models. This is especially important in the deep learning context due to the inherent complexity of these models and the challenges involved in interpreting their decision-making processes. There are several XAI techniques available to explain the decision of these models, i.e., CAM (Class Activation Mapping), Grad-CAM (Gradient-weighted Class Activation Mapping), SHAP (Shapley Additive Explanations), and LIME (Local Interpretable Model-agnostic Explanations), etc. \cite{li2022interpretable}. 

Among the previously mentioned XAI techniques, we adopt SHAP and LIME due to their model-agnostic nature. However, applying these methods individually to our model does not provide the level of explanation we require. Hence, we propose a novel ensemble approach that combines the strengths of both SHAP and LIME. 

Among the two available datasets, the IARC dataset is used for the types classification, while the AnnoCerv dataset is used for the CIN score classification. Performing explainability on the types and the CIN score is difficult, as we cannot verify the type and the CIN score with naked eyes. Hence, the easiest way to verify the explainability results is to apply explainability to the abnormal images, which are also available in the IARC dataset. Due to computational constraints, we work with about one-third of such abnormal images \cite{ ahamed2024detection}.

It is a known fact that cervical cancer typically occurs in the region around the cervix \cite{hemalatha2023cervixfuzzyfusion}. Hence, for the original abnormal images being studied, we identify the abnormality and mark this region with a digital pen. For the SHAP model, the output is in the form of two sets of images, the SHAP normal image and the SHAP abnormal image, with the region contributing to the decision-making of normal-abnormal highlighted in red color. If the marked cancerous region in the original abnormal image matches with the red region in the SHAP abnormal image, then we say it is a correct explanation. In all the other cases, we say the explanation is incorrect.

For the LIME technique, the output is only in the form of one image where the region that has contributed to the decision-making of normal-abnormal is present in the image, and the region that has not contributed is blackened. If the marked cancerous region in the original abnormal image matches with the un-blackened region in the LIME image, then we say that is a correct explanation. In all the other cases, we say the explanation is incorrect. Algorithm~\ref{algo:ensemble_xai} outlines the detailed step-by-step procedure of the proposed ensemble explainability method.

\begin{algorithm}[!htbp]
\caption{Ensemble Explainability Using SHAP and LIME}
\label{algo:ensemble_xai}
\begin{algorithmic}[1]
	\Require Set of abnormal images $\mathcal{I} = \{I_1, I_2, \dots, I_n\}$ with ground truth marked regions $\mathcal{R}^{\text{GT}}$
	\Ensure Explanation result (Correct or Incorrect) for each image
	
	\For{each image $I_k \in \mathcal{I}$}
	\State \textbf{Step 1: SHAP Explanation}
	\State \quad Generate SHAP images: $I_k^{\text{SHAP-normal}}$ and $I_k^{\text{SHAP-abnormal}}$
	\State \quad Extract red-highlighted region $\mathcal{R}_k^{\text{SHAP}}$ from $I_k^{\text{SHAP-abnormal}}$
	
	\State \textbf{Step 2: LIME Explanation}
	\State \quad Generate LIME explanation image $I_k^{\text{LIME}}$
	\State \quad Extract visible (non-black) region $\mathcal{R}_k^{\text{LIME}}$ from $I_k^{\text{LIME}}$
	
	\State \textbf{Step 3: Check Overlap with Ground Truth}
	\If{$\mathcal{R}_k^{\text{SHAP}}$ overlaps with $\mathcal{R}_k^{\text{GT}}$ \textbf{or} $\mathcal{R}_k^{\text{LIME}}$ overlaps with $\mathcal{R}_k^{\text{GT}}$}
	\State Mark explanation as \textbf{Correct}
	\Else
	\State Mark explanation as \textbf{Incorrect}
	\EndIf
	\EndFor
\end{algorithmic}
\end{algorithm}

The results of this approach for about one-third of available abnormal images ($122$) are presented in Table \ref{tab:XAI_results}. As shown in the table, the first row highlights the performance of the SHAP technique, which independently provides correct explanations for $96$ images and incorrect explanations for $26$ images, achieving a performance of $78.68\%$. The second row presents the performance of LIME, which independently generates the correct explanations for $92$ images and incorrect explanations for $30$ images, achieving a performance of $75.40\%$. 

\renewcommand{\arraystretch}{1.3}
\begin{table*}[!h]
\centering
\caption{Comparison of the XAI techniques for the abnormal images}
\vspace{5pt}
\resizebox{.85\textwidth}{!}{
	\begin{tabular}{c|c|c|c}
		\hline
		XAI  & No. of Correctly   &  No. of Incorrectly  & Performance  \\
		Techniques& Explained Images & Explained Images& (\%)\\\hline
		
		Independent SHAP &96 &26 & 78.68\\
		Independent LIME &92 & 30 & 75.40\\ 
		Collective Performance $\mathrm{I}$ &117 &5& 95.90 \\ 
		Collective Performance $\mathrm{II}$ & 119 & 3 & 97.54 \\\hline  
		
\end{tabular}}

\label{tab:XAI_results}
\end{table*}

\begin{figure*}[!h]
	\centering
	\begin{subfigure}[b]{0.52\linewidth}
		\centering
		\includegraphics[width=\linewidth]{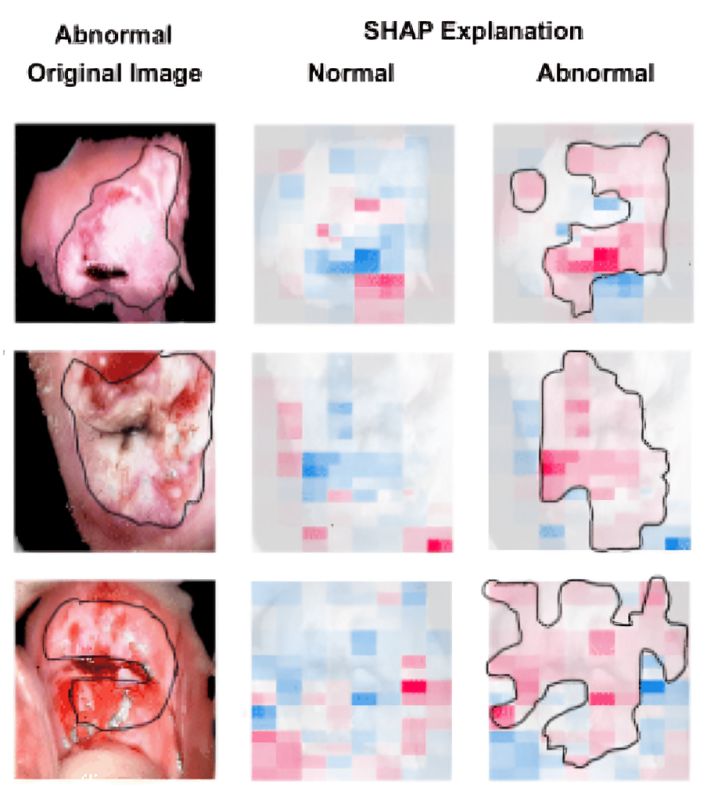}
		\caption{
		}
		\label{fig:correct_classified_1}
	\end{subfigure}
	\hfill
	\begin{subfigure}[b]{0.386\linewidth}
		\centering
		\includegraphics[width=\linewidth]{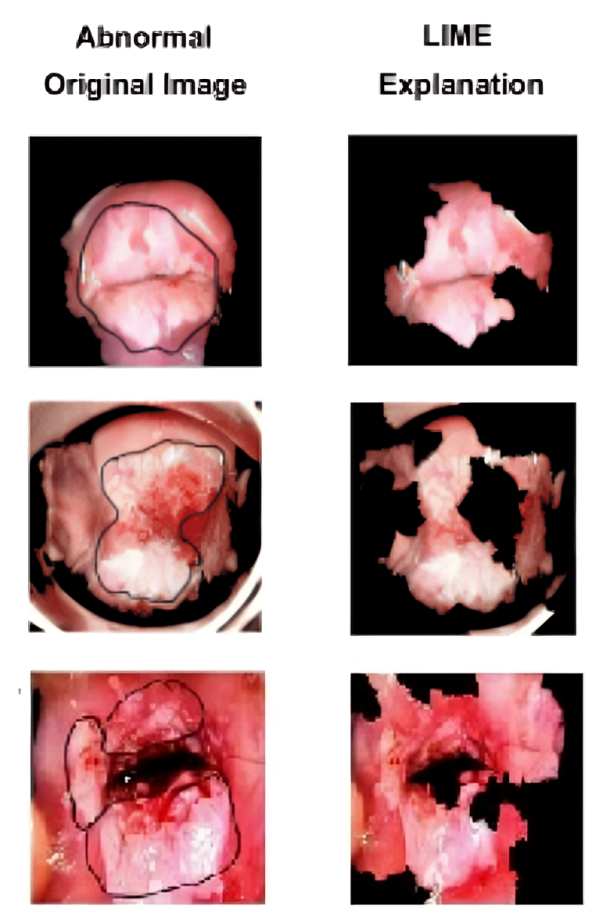}
		\caption{
		}
		\label{fig:correct_classified_2}
	\end{subfigure}
	\vspace{0.3cm}
	\caption{Examples of abnormal cervical images correctly explained using SHAP (a) and LIME (b) techniques, adapted from IARC dataset  \cite{IARC2024}.}
	\label{fig:correct_classified}
\end{figure*}

%
The third row of the table lists the data for the collective performance $\mathrm{I}$. Here, initially, LIME is applied, and then for those images where LIME fails, SHAP is applied. Thus, we achieve correct explanation for $117$ images ($92$ images by LIME and $25$ images by SHAP) and incorrect for $5$ images, achieving a performance of $95.90\%$. Similarly, the fourth row of the table lists the data for the collective performance $\mathrm{II}$. Here, initially, SHAP is applied, and then, for those images where SHAP fails, LIME is applied. Again, we achieve correct explanation for $119$ images ($96$ images by SHAP and $23$ images by LIME) and incorrect for $3$ images, achieving a performance of $97.54\%$.

For the sake of exposition, we present a subset of images. Fig.~\ref{fig:correct_classified_1} illustrates three cases where SHAP provides accurate explanations, while an additional eight images are included in Appendix~\ref{sec:appendix_A}. Similarly, Fig.~\ref{fig:correct_classified_2} shows three cases where LIME performs well, with eight more examples provided in Appendix~\ref{sec:appendix_B}. All $122$ abnormal images used in this analysis are available in the shared drive\footnote{\href{https://drive.google.com/drive/folders/1Ty-vWEh8zZgrGRBYSZG0tn1WgaRlyYGQ?usp=sharing}{Google Drive folder containing all 122 abnormal images with explaaination}}. Since the region contributing to the decision in our classification model is correctly identified by our two ensemble XAI techniques in most of the cases, we can confidently say that our classification model performs correct classification.

\section{Conclusions and future work}\label{sec:conclusion}

We propose a CAD system for better cervical cancer image classification. ResNets are employed as the feature extraction model in this work, due to their recent success in medical imaging and their ability to address the vanishing gradient problem. Since lower and higher numbered ResNets capture simple and complex patterns, respectively, we use three such models, i.e., ResNet50, ResNet101, and ResNet152.

In this work, we customize the above-listed ResNets and develop novel feature extraction models that have three essential modules. \textit{First}, we extract the detailed features from the earlier convolution blocks and the abstract features from the later convolution blocks, because all kinds of features are important for cervical cancer classification. This is first attempt of its kind in any type cancer classification. \textit{Second}, these extracted features are enhanced by a non-parametric 3D attention module, which is uniquely incorporated into each convolution block of our model. \textit{Third}, these enhanced features are passed through our innovative adaptive pooling module for feature selection as they are high-dimensional. Here, the detailed features which are noise-prone are passed through the GMP layer, and the noise-resilient abstract features are passed through the GAP layer. This process results in three new models, termed BF-AD-AP-ResNet50, BF-AD-AP-ResNet101, and BF-AD-AP-ResNet152. 

To achieve a more comprehensive and informative feature space, we design a Tri-Stream model, which unifies the attention-enhanced features derived from our Proposed BF-AD-AP-ResNets models. The resulting feature representation is classified using SVM classifier.

We perform our experiments on two datasets, namely, the IARC dataset provided by WHO and the AnnoCerv dataset. On the IARC dataset, the standard ResNets achieves an average performance of $90.91\%$, while our best model achieves a significantly higher average performance of $\textbf{98.63\%}$. Compared to the best available competitive approach on this dataset, our method shows an average improvement of $\textbf{14.55\%}$. On the AnnoCerv dataset, the standard ResNets attain an average performance of $87.68\%$, whereas our best model achieves an improved average performance of $\textbf{93.39\%}$. To the best of our knowledge, there is currently no competitive approach available for comparison on this dataset.

We carry out ablation study of each component of our model because it allows us to systematically evaluate the individual contribution and importance of every module, ensuring that each part genuinely enhances the overall performance rather than adding redundant complexity.

We also study the explainability of our classification models where we propose a new ensemble of SHAP and LIME XAI techniques. It is a well-known fact that cervical cancer originates around the cervix region. The region that contributes to the decision-making in our classification models, as identified by our ensemble of XAI techniques, turns out to be around the cervix in $\textbf{97\%}$ of the cases. Hence, this validates the correctness of our classification models.  

One of the future work directions is to test our CAD system on a much larger dataset, as WHO has mentioned that they would provide such a dataset soon. A second direction for future work is to study the implicit relation between different components \cite{kim2005effectiveness}. Another future work could be to formulate the combination of different deep neural networks as an optimization problem as done in different domains \cite{agarwal2019Paralarpd, ahuja2008Mixed}. Finally, it would be good to explore the possibility of approximate computing \cite{gupta2020ApproxProc, gupta2020Approx} in this context.

\bmhead*{Acknowledgements}
We would like to thank Mr. Siddartha Chennareddy and Mr. Karthik Boddupalli for their valuable discussions on various aspects of this research.

\section*{Statements \& Declarations}

\bmhead*{Authors Contributions} 
Saurabh Saini performed investigation, data curation, and writing - original draft. Kapil Ahuja conceptualised, performed project administration, and writing - review \& editing. Akshat S. Chauhan performed formal analysis and software preparation.


\bmhead*{Funding} This study was not funded. The authors have no relevant financial or non-financial interests to disclose.


\bmhead*{Data availability} 
The IARC colposcopy dataset can be obtained upon request \cite{IARC2024}, while the AnnoCerv dataset is available in \cite{Minciuna2023}.

\bmhead*{Competing Interest} The authors state they have no financial or personal conflicts related to this research, authorship, or publication.

\bmhead*{Ethical Approval} No experiments involving human participants or animals were conducted by the authors in this study.

\begin{appendices}
	
	\section{XAI Results (SHAP)}\label{sec:appendix_A}
	\begin{figure*}[!htbp]
		\centering
		\includegraphics[width=\linewidth]{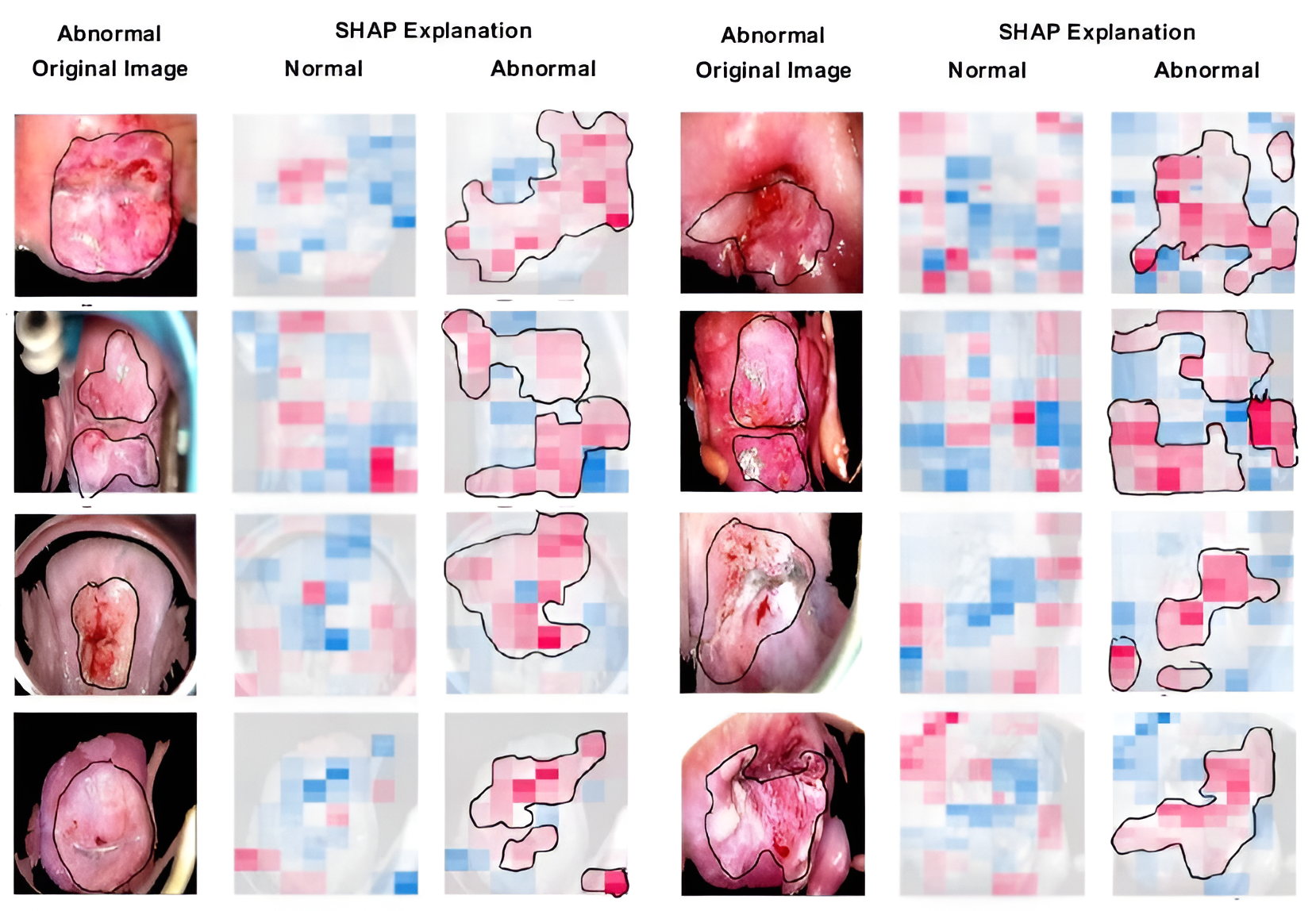}
		\vspace{0.3cm}
		\caption{Examples of abnormal cervical images correctly explained using the SHAP technique, adapted from IARC dataset \cite{IARC2024}}
		\label{fig:correct_classified_3}
	\end{figure*}
	\FloatBarrier
	\pagebreak
	\section{XAI Results (LIME)}\label{sec:appendix_B}
	\begin{figure*}[!htbp]
		\centering
		\includegraphics[width=0.85\linewidth]{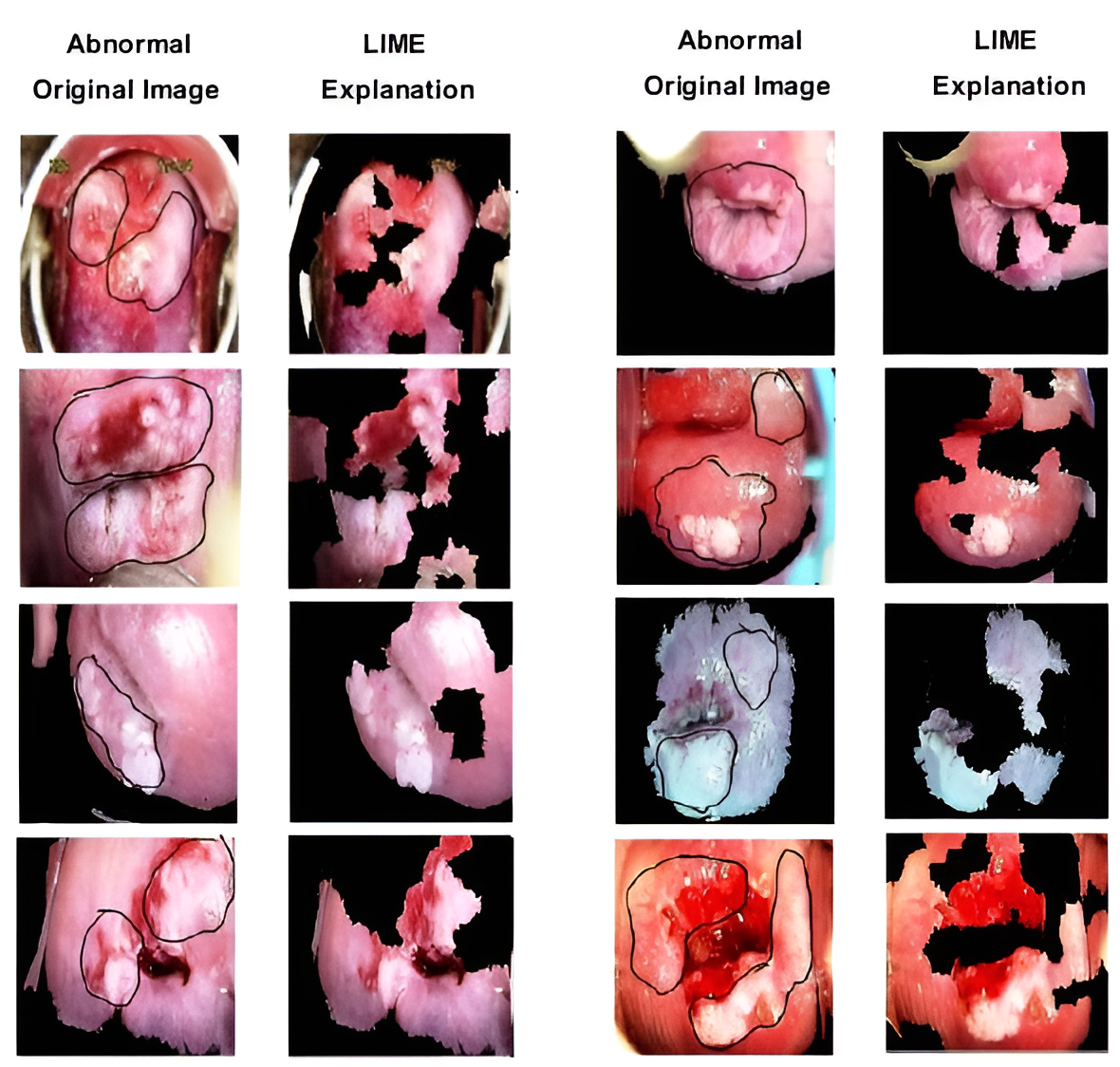}
		\vspace{0.3cm}
		\caption{Examples of abnormal cervical images correctly explained using the LIME technique, adapted from IARC dataset \cite{IARC2024}}
		\label{fig:correct_classified_4}
	\end{figure*}
	\FloatBarrier
\end{appendices}

	\backmatter
\maketitle

	\bibliography{sn-bibliography}

\end{document}